\documentclass{article}

\usepackage{arxiv}
\usepackage[utf8]{inputenc} 
\usepackage[T1]{fontenc}    
\usepackage{hyperref}       
\usepackage{url}            
\usepackage{booktabs}      
\usepackage{amsfonts}       
\usepackage{nicefrac}       
\usepackage{microtype}      
\usepackage{lipsum}
\usepackage{graphicx}
\usepackage{tabularx}
\usepackage{booktabs}  
\usepackage{array}    
\usepackage[normalem]{ulem}
\usepackage{algorithm}
\usepackage{algpseudocode} 
\usepackage{amsmath}
\usepackage{subcaption}
\usepackage[table,xcdraw]{xcolor}
\usepackage{colortbl}
\usepackage{pifont}
\usepackage[labelfont=it]{caption}
\useunder{\uline}{\ul}{}
\graphicspath{ {./images/} }

\title{Detecting Reading-Induced Confusion Using EEG and Eye Tracking}

\author{
Haojun Zhuang\textsuperscript{1},
Dünya Baradari\textsuperscript{2},
Nataliya Kosmyna\textsuperscript{2},
Arnav Balyan\textsuperscript{2} \\
\textbf{Constanze Albrecht}\textsuperscript{2},
\textbf{Stephanie Chen}\textsuperscript{3},
\textbf{Pattie Maes}\textsuperscript{2} \\
\\
\textsuperscript{1}University of California, Berkeley, USA \\
\textsuperscript{2}MIT Media Lab, USA \\
\textsuperscript{3}Princeton University, USA \\
}

\begin{document}
\maketitle
\begin{abstract}
Humans regularly navigate an overwhelming amount of information via text media, whether reading articles, browsing social media, or interacting with chatbots. Confusion naturally arises when new information conflicts with or exceeds a reader’s comprehension or prior knowledge, posing a challenge for learning. In this study, we present a multimodal investigation of reading-induced confusion using EEG and eye tracking. We collected neural and gaze data from 11 adult participants as they read short paragraphs sampled from diverse, real-world sources. By isolating the N400 event-related potential (ERP), a well-established neural marker of semantic incongruence, and integrating behavioral markers from eye tracking, we provide a detailed analysis of the neural and behavioral correlates of confusion during naturalistic reading. Using machine learning, we show that multimodal (EEG + eye tracking) models improve classification accuracy by 4-22\% over unimodal baselines, reaching an average weighted participant accuracy of 77.3\% and a best accuracy of 89.6\%. Our results highlight the dominance of the brain's temporal regions in these neural signatures of confusion, suggesting avenues for wearable, low-electrode brain-computer interfaces (BCI) for real-time monitoring. These findings lay the foundation for developing adaptive systems that dynamically detect and respond to user confusion, with potential applications in personalized learning, human-computer interaction, and accessibility.
\end{abstract}

\section{Introduction}
Confusion is a critical cognitive-affective state that arises when individuals encounter information that is ambiguous, contradictory, or difficult to integrate \cite{d2012dynamics}. This state is closely linked to cognitive disequilibrium or dissonance, in which a person's internal understanding clashes with external information \cite{robertson2015international}. Although typically considered uncomfortable, confusion facilitates learning, decision-making, and comprehension \cite{d2012dynamics}. Cognitive theories attribute their occurrence to conflicts between fast, heuristic-based processing (Type 1) and slower, analytical reasoning systems (Type 2) \cite{kahneman2011thinking, stanovich2000individual}. When these systems diverge, confusion can trigger deliberate, effortful cognitive activity to resolve discrepancies, which in turn supports deeper understanding and integration of new knowledge \cite{d2012dynamics}. Empirical research supports this, showing that learners who experienced confusion during study sessions demonstrate significantly greater learning gains. For instance, one study reported a 46\% increase in learning compared to those who did not experience confusion \cite{craig2004affect}. Furthermore, recent advances in adaptive learning systems underscore the value of real-time detection and resolution of confusion, especially for individuals with cognitive impairments or language barriers, emphasizing its central role in adaptive interface design \cite{sadras2023post}.

However, reliably measuring confusion states remains challenging. Traditional methods for confusion detection often rely on subjective self-reports, such as questionnaires like the Self-Assessment Manikin (SAM) \cite{morris1995observations}. While these approaches provide valuable insights, they lack the granularity and scalability required for real-world applications. Existing research in neuroscience and cognitive science has explored physiological markers of confusion, utilizing modalities such as eye tracking and electroencephalography (EEG). A key electrophysiological marker associated with stimulus congruence is the N400 event-related potential (ERP), a negative deflection in EEG signals peaking around 400 ms after stimulus presentation \cite{kutas2011thirty}. N400 has been used extensively to study semantic incongruities \cite{frishkoff2004frontal, desai2023real}, linguistic anomalies \cite{osterhout2002brain}, and contextual violations \cite{kutas1980reading, he2024multivariate, rabs2022situational}, providing a neural basis for detecting confusion. However, many text-based N400 ERP studies have employed highly controlled word-by-word reading paradigms, in which only a single word is presented at a time to minimize eye movement artifacts. Other work has utilized alternative modalities such as audio \cite{kallionpaa2019single} or video \cite{wang2013using,kopparapu2023spatial,li2019n400,ni2017confused,trigka2023mental,chowdhuri2024detection}. While these approaches effectively control for experimental artifacts, they suffer from low ecological validity, as they fail to capture the natural flow and complexity of real-world reading, where individuals typically process full sentences or paragraphs rather than isolated words.

While confusion has been examined in domains such as conceptual learning, problem-solving, and digital tutoring systems, the specific dynamics of confusion that arise under natural reading, particularly semantic or pragmatic dissonance within continuous text, remain underexplored. This distinction is essential, as reading-induced confusion is often transient, covert, and sensitive to fine-grained linguistic cues, necessitating detection strategies that differ from those used in task-based or performance-oriented environments.

Addressing this gap, our study investigates the neural and behavioral signatures of confusion as they naturally occur during paragraph reading, leveraging multimodal EEG and eye tracking data. We specifically examine reading-induced confusion, which we define as the cognitive response that arises during the visual processing of well-formed but conceptually challenging text, not merely as a reaction to linguistic anomalies or syntactic violations. Our work builds on prior research demonstrating that N400 responses are modulated not only by semantic association but also by event-level expectancy \cite{rabs2022situational}. Such findings highlight that even semantically related words can elicit pronounced N400 responses if they are implausible within the broader situational context. This reinforces our interpretation of the N400 as a marker of confusion rooted in conceptual conflict rather than solely lexical incongruity. 

To detect confusion, we conducted a multimodal experiment collecting EEG and eye tracking data from 11 participants. Our experiment extends on prior work by examining confusion elicited by short text paragraphs collected from diverse, real-world sources, such as Wikipedia articles and Amazon reviews. We curated these paragraphs to induce two subtypes of confusion: (1) \textit{Factual Confusion}, arising from contradictions to previously acquired knowledge, and (2) \textit{Contextual Confusion}, stemming from insufficient background knowledge (Table \ref{tab:table-1}). This distinction acknowledges the diversity of confusion that can occur during reading. By clearly defining and isolating these two subtypes, we enhance the interpretability and robustness of our findings and enable more precise comparisons between confusion subclasses. Additionally, we included easily comprehensible paragraphs as a control condition.

Our key contributions are as follows:
\begin{enumerate}
    \item \textit{Text corpus and multimodal dataset for confusion induction}: We present a novel corpus of paragraphs drawn from diverse real-world contexts, designed to elicit \textit{Factual} or \textit{Contextual Confusion} during reading, accompanied by non-confusing control paragraphs. We provide a dataset comprising EEG and eye-tracking measurements from 11 participants, as they each read 300 paragraphs.
    \item \textit{Confusion Classification pipeline}: Using the dataset, we developed a machine learning pipeline that achieves an average weighted test accuracy of 77.29\% and a best test accuracy of 89.55\% in identifying confusion states.
    \item \textit{Exploration of N400 and eye fixation markers}: We examine neural and behavioral indicators of confusion, including the N400 event-related potential and gaze-based markers during paragraph-level reading tasks.
    \item \textit{Neurophysiological insights}: Our findings highlight the temporal region’s dominance in reading-induced confusion, suggesting potential avenues for wearable, low-electrode brain-computer interface (BCI) applications.
\end{enumerate}

\section{Background and Related Work}

\subsection{Confusion Detection using EEG}

Electroencephalography (EEG) is a non-invasive neuroimaging technique that captures the electrical activity of the brain by measuring the synchronized firing of large groups of neurons via electrodes placed on the scalp \cite{niedermeyer2005electroencephalography,nunez2006electric}. This voltage signal reflects rhythmic fluctuations within distinct frequency bands, such as delta, theta, alpha, beta, and gamma, each associated with specific brain states. EEG has been widely used to assess cognitive processes such as attention \cite{sun2025attention,kaushik2022decoding,liu2013recognizing}, cognitive load \cite{zhu2021study,antonenko2010using,friedman2019eeg}, decision-making \cite{gao2023neural,si2020predicting,li2025neural}, and engagement \cite{apicella2022eeg,berka2007eeg,pope1995biocybernetic}, which are critical for learning and performance. It has been applied to track mental effort in problem-solving tasks \cite{zhu2021study}, measure cognitive load levels during multitasking \cite{lim2018stew}
, or monitor attention in real-world environments \cite{kaushik2022decoding}. Furthermore, EEG has been used to study engagement in learning \cite{apicella2022eeg,berka2007eeg} and develop biofeedback systems for personalized education \cite{kosmyna2019attentivu,liu2013recognizing,mills2017put}. Its high temporal resolution makes EEG particularly suitable for detecting confusion, a transient cognitive-affective state characterized by uncertainty, mental effort, or conflict in processing information, often preceding learning or disengagement. Consequently, EEG serves as an excellent tool for adaptive learning and human-computer interaction (HCI) systems, where insights into cognitive states can significantly enhance user experience and inform personalized feedback strategies.

EEG studies investigating confusion have explored various modalities to induce states of confusion, including text, audio \cite{kallionpaa2019single,li2019n400}, images \cite{dini2022eeg}, video \cite{wang2013using,li2019n400}, adventure games \cite{benlamine2021confusion},  reasoning and problem-solving tasks \cite{xu2023confused,zhou2018confusion,dakoure2021confusion,mampusti2011measuring}. However, standards or unified methods for detecting confusion do not exist, and research on affective markers of confusion is still sparse and fragmented \cite{xu2023confused,ganepola2024systematic}. Measuring confusion, therefore, often relies on self-assessment; asking participants to rate their level of confusion using binary or Likert-type scales \cite{benlamine2021confusion,dakoure2021confusion,mampusti2011measuring} or questionnaires such as the standardized Self-Assessment Manikin (SAM) \cite{morris1995observations}. The physiological markers are then correlated with the self-assessment. 

A recent systematic review by Ganepola et al. \cite{ganepola2024systematic} found that despite increasing interest in EEG-based emotion recognition, only five studies to date have collected their own EEG datasets for confusion. Of these, only one is publicly available, underscoring the scarcity of reliable labeled data in this area. The review also emphasized a lack of standardization across experimental protocols, EEG preprocessing methods, and emotion labeling criteria, all of which hinder cross-study comparability and reproducibility. Ganepola et al. grouped confusion detection approaches into three categories: those relying solely on self-report, those integrating behavioral or performance metrics, and hybrid methods combining multimodal signals. Machine learning models ranging from traditional classifiers such as Support Vector Machines (SVMs), k-Nearest Neighbors (k-NNs), and Random Forests to deep learning approaches like Convolutional Neural Networks (CNNs) have shown promise in decoding confusion from EEG features. These models are particularly effective when incorporating both spatial and spectral information from multiple channels. Nevertheless, the authors emphasize that generalizability remains a significant challenge, largely due to inter-subject variability and small sample sizes.

Although the literature presents a range of methods, several studies are particularly notable for their experimental designs that introduce more control or richer labeling strategies beyond simple self-assessment.

A notable study by Wang et al. \cite{wang_confused_eeg_2016} leveraged videos from Massive Open Online Courses (MOOCs) \cite{wang2013using} to explore confusion detection using EEG. The authors curated two sets of educational videos: one containing basic topics such as elementary mathematics, which were expected to be easily understood by college students, and another featuring complex subjects like quantum mechanics and stem cell biology, which required substantial prior knowledge. To further induce confusion, content from each two-minute video was deliberately removed in the middle. After watching each clip, participants self-rated their level of confusion, which was treated as the ground truth, while EEG data was recorded throughout. In addition, three external observers assessed participants’ confusion based on body language and facial expressions. The study showed that machine learning models trained on EEG data could predict confusion levels with an accuracy comparable to that of human observers. The dataset was made publicly available \cite{wang_confused_eeg_2016}, enabling subsequent research that has further advanced confusion classification using various machine learning approaches \cite{kopparapu2023spatial,ni2017confused,trigka2023mental,chowdhuri2024detection}. However, a key limitation of the original study is its use of a single-channel EEG system, which severely limits the spatial resolution of the recordings. Ganepola et al. \cite{ganepola2024systematic} identified this dataset as one of the most influential in the field due to its accessibility and emphasized the need for future work to adopt high-density EEG setups to enhance both spatial and spectral feature resolution.

Other studies infer confusion based on task performance or quiz results, an approach common in problem-solving tasks \cite{dakoure2021confusion,mampusti2011measuring}. For example, Dakoure et al. \cite{dakoure2021confusion} collected EEG data as participants completed five cognitive ability tests, including matrix reasoning, verbal analogies, maze solving, and memory tasks, while using self-reports to classify confusion. Zhou et al. \cite{zhou2018confusion} implemented an elegant experimental setup in which participants completed a series of Raven’s Progressive Matrices (RPM), a widely recognized psychological assessment for evaluating abstract reasoning abilities. Half of the RPM items were intentionally designed to be unsolvable within 15 seconds, thereby inducing confusion. In a subsequent study, some of the same authors extended the task to the more challenging Advanced Progressive Matrices (APM), and combined self-reported confusion with objective task performance \cite{xu2023confused}. Responses were classified as 'Non-Confused' if the participant solved a matrix correctly without reporting confusion, or as a 'Guess' if they solved it correctly but felt confused. Incorrect answers paired with self-reported confusion were labeled ‘Confused,’ while incorrect but confident responses were categorized as ‘Think-Right.’ By distinguishing between ‘Guesses’ and ‘Think-Right’ responses, the researchers refined their confusion labeling for greater accuracy. These types of study designs that combine performance outcomes with self-report are considered the most robust for eliciting and measuring confusion, although their task-specific nature can limit model generalizability across domains \cite{ganepola2024systematic}. However, not all experimental paradigms include tasks where performance can be objectively measured and directly compared to self-reported confusion.

\subsection{N400 and Semantic Processing}

For language understanding, the N400 event-related potential (ERP) is one of the most widely used EEG measures that approximates confusion. An event-related potential is a brain response that is time-locked to a specific sensory, cognitive, or motor event \cite{penny2002event,kosmyna2019conceptual}. The N400, in particular, is recognized as a neural marker of semantic processing and prediction error. It appears as a negative deflection in the EEG signal, peaking approximately 400 milliseconds after the presentation of a textual, visual, or auditory stimulus that is semantically incongruent or unexpected \cite{vsovskic2022better,kutas1984brain,kutas1980reading,lau2016direct}. For example, in the sentence \textit{"For breakfast, I ate a bagel and a giraffe,"} the word \textit{"giraffe"} is contextually unexpected and reliably elicits a strong N400 response. 

The N400 is also considered a neural marker of predictive coding during language comprehension \cite{kutas2011thirty,kutas1980reading,heikel2018decoding}. In psycholinguistics, Cloze probability is commonly used to quantify how predictable a word is within its context and to investigate semantic processing \cite{michaelov2022so}. Cloze probability is defined as the proportion of people who complete a sentence fragment with a particular word, typically measured by asking participants to provide likely sentence completions. Cloze probabilities range from 0 (no one predicts the next word) to 1 (everyone predicts it). Crucially, the N400 amplitude is inversely related to semantic predictability: the less expected a word is, the larger (more negative) the N400 response, and vice versa \cite{kutas1984brain}. This relationship has led to the hypothesis that the N400 may index a neural representation of semantic space, analogous to word embeddings in natural language processing (NLP) and large language models (LLMs). Supporting this idea, recent studies have demonstrated that EEG signals recorded during isolated word reading both reflect and predict a word’s position in distributional semantic space (e.g., word2vec), particularly in the 300–500 ms post-stimulus window \cite{sassenhagen2020traces}.

N400 responses have also been elicited across multiple modalities, reinforcing their value as a proxy for confusion beyond just traditional linguistic contexts. While originally identified in response to semantic anomalies in text, subsequent research demonstrates that N400 effects also arise in visual, auditory \cite{kallionpaa2019single,heikel2018decoding}, and cross-modal paradigms \cite{li2019n400}. For example, Kellenbach et al. \cite{kellenbach2000visual} showed that visual-perceptual features influence semantic processing by presenting participants with word pairs that are either visually similar (e.g., button–coin) or dissimilar. Visually similar pairs elicited a reduced N400 component, indicating that perceptual attributes contribute to semantic priming. More recently, Dini et al. \cite{dini2022eeg} investigated N400 responses in a branding context, where participants viewed brand-related image sets that were either congruent (brand-related images matched the brand logo) or incongruent (brand-related images mismatched the brand logo). EEG recordings revealed that incongruent brand logos elicited a significantly larger N400 deflection and increased theta power compared to congruent ones. Together, these findings underscore the N400’s sensitivity to both linguistic and perceptual mismatches, supporting its broader role in conceptual processing across modalities.

Despite the widespread use of the N400 component in detecting confusion and semantic incongruence, previous studies face certain challenges in capturing the complexity of real-world semantic processing: 

\begin {enumerate}
    \item To elicit a strong N400 response, experimenters often artificially engineer sentences by inserting semantically incongruent or unpredictable words (e.g., \textit{"She spread butter on her freshly toasted sock."}), as seen in seminal studies such as Kutas and Hillyard \cite{kutas1980reading} and subsequent reliability studies \cite{cruse2014reliability}. While initial sentence structures are often sourced from corpora such as novels \cite{frank2015erp}, researchers typically replace words with semantically implausible target words, resulting in expectancy violations that rarely occur in natural text. Some studies use adjective-noun paradigms to precisely control predictability via corpus statistics \cite{lau2016direct}. However, these approaches still rely on artificially constructed violations rather than naturally occurring ambiguities in authentic reading.
    \item Many experiments employ contextually isolated stimuli, such as individual sentences, rather than coherent paragraphs that better represent natural reading. Some recent work, such as Rabs et al. \cite{rabs2022situational}, has begun to address this by using multi-sentence discourse contexts, moving stimulus design closer to real-world reading scenarios.
    \item A common constraint in traditional text-based N400 experiments is the reliance on word-by-word reading paradigms. To minimize eye movement artifacts in EEG, most studies present text stimuli incrementally, one word at a time, on a computer screen, trading the natural flow of reading for cleaner data. This methodological choice, while effective for artifact reduction, limits ecological validity and disrupts authentic comprehension processes. Given these constraints, there is a growing need for experimental paradigms that better capture real-world language comprehension.
\end {enumerate}

\subsection{Eye Tracking for Reading-Based Confusion Detection}

Eye tracking is commonly used to detect affective and cognitive states in users. Most prior studies have focused on emotionally evocative video stimuli (e.g., to induce joy or sadness), often using pupil dilation as the primary measure \cite{alghowinem2014exploring,liu2019research,alshehri2013exploratory}. However, research leveraging eye tracking with textual stimuli remains limited. While pupil dilation is effective for capturing arousal during multimedia content, it offers limited temporal precision in reading tasks, where short fixations and rapid saccades constrain its ability to resolve transient cognitive states such as confusion. In the context of reading, \textit{fixation}, defined as a sustained gaze on a specific visual region, serves as a primary metric for assessing cognitive processing. Eye-tracking technologies can capture critical measures, including fixation duration, pupil dilation, and gaze entropy, all of which reflect cognitive effort and attention shifts during reading comprehension \cite{shiferaw2019review}. Recent work has shown that gaze patterns during the reading of conflicting information, particularly regressions between competing claims and evidence, can reflect strategic evaluation efforts and early cognitive disequilibrium, i.e., confusion \cite{tsai2022critical}. These patterns are especially pronounced in highly critical readers, who tend to fixate more and revisit reasoning- and evidence-rich sections. Moreover, fixations on irrelevant or extraneous content have been correlated with elevated self-rated confusion, suggesting that confusion may manifest in gaze behavior even before readers become consciously aware of it or performance begins to degrade \cite{pachman2016eye}.

Some studies have exclusively used eye tracking to examine cognitive states related to confusion. For instance, Sims et al. \cite{sims2020neural} applied RNN and CNN architectures on eye tracking data, achieving a high accuracy of 79\% in detecting user confusion during interaction with the decision-support tool ValueChart. Similarly, Child et al. \cite{child2020tracking} found prolonged fixation durations when participants read texts from third-person perspectives compared to second-person perspectives. Increased fixation frequencies have also been observed in response to negative emotional valence in text, as opposed to positive valence \cite{arfe2023effects}. Importantly, the predictive utility of eye tracking extends beyond content-level features: metrics such as regression rate, gaze duration, and first fixation timing collectively explain more variance in comprehension outcomes than reading speed alone, highlighting their potential in modeling latent reader states \cite{meziere2023using}.

\subsection{Eye-Tracking as a Complementary Modality}

Eye movements are intrinsically linked to cognitive processing, yet their presence in EEG studies introduces noise artifacts that complicate the analysis of neural data (typically mitigated by techniques such as Independent Component Analysis (ICA)) \cite{bigdely2015prep}. Traditional language comprehension paradigms attempt to mitigate this by presenting words one at a time, thereby minimizing eye movement. This approach, however, substantially reduces ecological validity and disrupts natural reading dynamics, restricting cognitive research to simplified and artificial contexts.

To overcome these limitations, recent research has increasingly adopted multimodal approaches that combine EEG with eye tracking and electrooculography (EOG), which captures electrical ocular activity. Such integrations not only manage ocular artifacts but also enable the study of richer, continuous text stimuli, thereby enhancing ecological validity \cite{delogu2017teasing,white2009wait,chen2015automatic,hollenstein2021decoding}. Notably, Delogu et al. \cite{delogu2017teasing} demonstrated that both neural (N400) and eye movement (fixation) measures are sensitive to semantic unpredictability and conceptual integration difficulty during sentence comprehension, such as that arising from epistemic ambiguity, conflicting claims, or dense discourse. Their findings support the value of multimodal approaches for detecting cognitive conflict and confusion during reading.

Benchmarks like EEGEyeNet exemplify the power of multimodal frameworks, achieving accurate real-time gaze estimation from EEG data and improved cognitive state classification \cite{kastrati2021eegeyenet}. Similarly, multimodal procedures have enabled enhanced semantic prediction and user state modeling \cite{jamal2023integration}, with studies like Zheng et al. \cite{zheng2014multimodal} showing gains in emotion detection accuracy by combining EEG and eye tracking. The Zurich Cognitive Language Processing Corpus (ZuCo) \cite{hollenstein2018zuco,hollenstein2019zuco} - and its extension, ZuCo 2.0 \cite{hollenstein2019zuco}) - stand out as the only dataset to date that aligns EEG and eye tracking data during paragraph-level, naturalistic reading. However, ZuCo was not designed to systematically elicit or annotate confusion; its textual materials were not curated to provoke cognitive conflict or ambiguity. Our work extends this multimodal, paragraph-level paradigm, but explicitly targets the detection and classification of confusion states through specifically crafted and categorized reading materials.

Despite these methodological advances, a clear gap remains: no dataset currently exists that is specifically designed to capture and classify confusion states during paragraph-based reading using multimodal technologies. Our study addresses this gap by explicitly focusing on the neural and ocular correlates of distinct confusion subtypes, under naturalistic yet carefully controlled conditions. In doing so, we aim to advance the understanding of cognitive processing and confusion dynamics during real-world reading, and provide a foundation for future adaptive systems that can sense and respond to user confusion in situ.


\section{System Design}

\begin{figure}[ht]
    \centering
    \includegraphics[width=\linewidth]{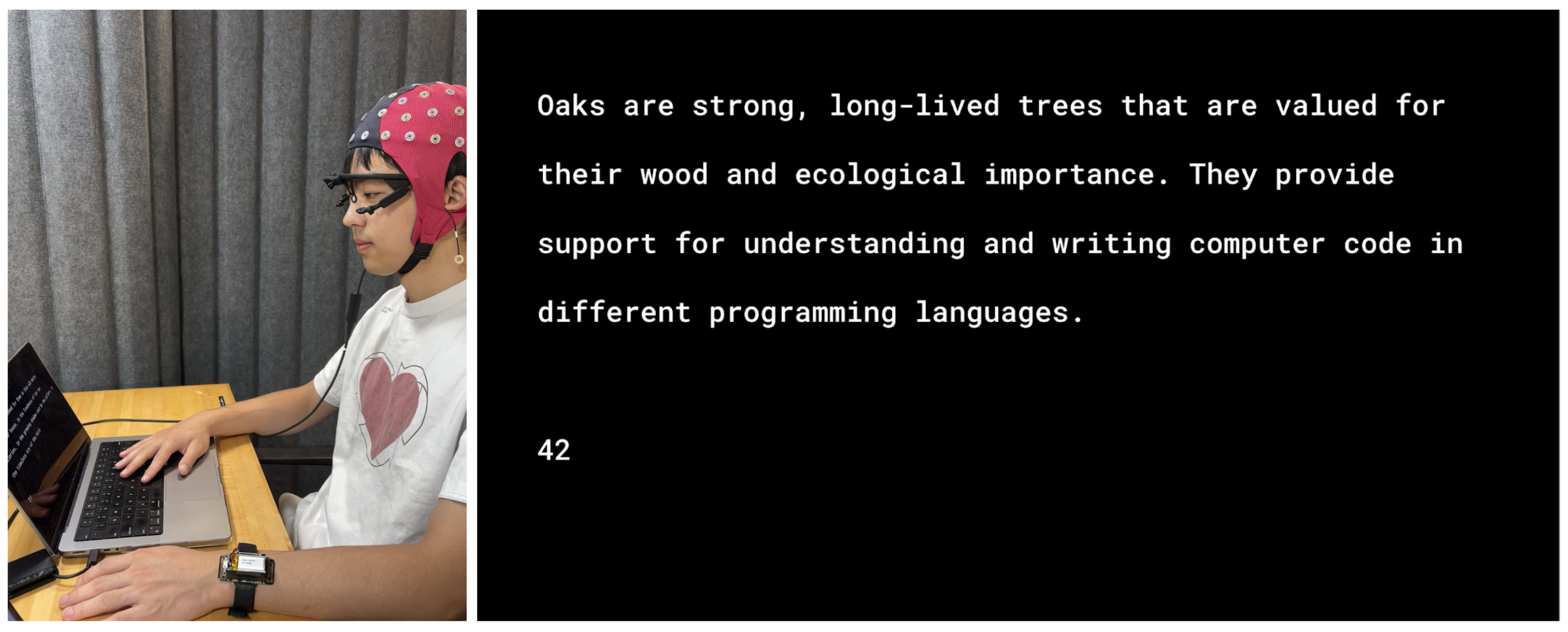}
    \caption{\textit{Left}: Participant wearing an EEG headset and eye tracking glasses while sitting in a dark, quiet room reading the text on a laptop screen. Photo taken for demonstration purposes in a well-lit room. \textit{Right}: Sample trial of the experiment (trial 42). White text is shown on a dark background for 10 seconds. In this case, the paragraph belongs to the \textit{Factual Confusion} class.
}
    \label{fig:fig1}
\end{figure}

\subsection{Text Curation}
We compiled a reading dataset of 300 paragraphs (with an average word count of 26.38/paragraph) from various sources, including Wikipedia, Amazon reviews, textbooks, and literary datasets (Table \ref{tab:table-2}). These diverse sources were chosen to resemble the real-life experience of reading in various contexts. We curated 120 \textit{Control} paragraphs, 99 \textit{Factual Confusion} paragraphs, and 81 \textit{Contextual Confusion} paragraphs as per our definitions, provided in Table \ref{tab:table-1}. The \textit{Factual Confusion} paragraphs were created by shuffling phrases/sentences around well-known facts. The \textit{Contextual Confusion} paragraphs were extracted from four domain-specific sources: medicine, machine learning, philosophy, and literature (see Table \ref{tab:table-2} for a detailed description).

\begin{table}[ht]
\caption{Overview of Confusion Classes with Examples}
\label{tab:table-1}
\renewcommand{\arraystretch}{1.5}
\begin{tabularx}{\textwidth}{lXXl}
\hline
\textbf{Class} & \textbf{Definition} & \textbf{Example Text} & \textbf{Number of Trials} \\
\hline
\textit{Control} &
Text is correct, easily comprehensible, and requires minimal context to understand. &
Polar bears are native to the Arctic and are the largest land carnivores. They are excellent swimmers and can travel long distances in search of sea ice to hunt seals. &
120 \\
\textit{Factual Confusion} &
Text is easily comprehensible but contradicts common knowledge. &
The Golden Gate Bridge is a seawater inlet of the Indian Ocean, and is located between Africa and Asia. &
99 \\
\textit{Contextual Confusion} &
Text is correct, but requires substantial prior knowledge of context to understand. &
The strength of the constraint is dictated by the neighborhood size. The larger the size of the neighborhood, the stronger the constraint, and the more sensitive the solution is to the particular choice of constraint. &
81 \\
\hline
\end{tabularx}
\end{table}

\begin{table}[ht]
\caption{Text Sources and Example Paragraphs}
\label{tab:table-2}
\small
\renewcommand{\arraystretch}{1.2}
\begin{tabularx}{\textwidth}{p{1.5cm}>{\itshape}p{1.5cm}|p{6cm}X}
\hline
\textbf{Source} & \textbf{Class Label} & \textbf{Note} & \textbf{Example Paragraph} \\
\hline
Facts &
  Control &
  The facts were manually compiled and checked by the authors. &
  Polar bears are native to the Arctic and are the largest land carnivores. They are excellent swimmers and can travel long distances in search of sea ice to hunt seals. \\
Wikipedia \cite{Cohere59} &
  Control &
  The first few sentences of Wikipedia entries. &
  Diana, Princess of Wales, was a member of the British royal family. She was the first wife of King Charles III (then Prince of Wales) and mother of Princes William and Harry. \\
Amazon \cite{McAuleyLab60} &
  Control &
  We only selected texts that could be understood without context. &
  Absolutely one of the most beautiful little necklaces around. Get it and you will get lots of compliments on how nice it is. I was so happy that I purchased it. I had been looking for over three years for the perfect one, and this is it. \\
Wikipedia \cite{Cohere59} &
  Factual Confusion &
  Created by falsely combining keywords. &
  Marilyn Monroe is a 2022 American superhero film based on Marvel Comics featuring the character Thor. \\
Facts &
  Factual Confusion &
  Created by falsely combining keywords. &
  Wolves are social animals and are known to live and hunt in packs. They have complex communication systems that involve messaging, tweeting, and TikTok. \\
Stanford Encyclopedia of Philosophy \cite{Ruggsea61} &
  Contextual Confusion &
  The Stanford Encyclopedia of Philosophy unites global scholars to maintain a current and authoritative reference in philosophy and related areas. &
  With the emergence of liberal democracy in the modern West, however, the types of questions that philosophers asked about the interrelation between religion and political authority began to shift. \\
Mayo Clinic Diseases \& Conditions \cite{MayoClinic62} &
  Contextual Confusion &
  An online knowledge platform providing easy-to-understand answers about diseases and conditions maintained by the Mayo Clinic.&
  The location of the median arcuate ligament and celiac artery varies slightly from person to person. Typically, the ligament runs across the largest blood vessel in the body (aorta). \\
Machine Learning Textbook \cite{hastie2017elements} &
  Contextual Confusion &
  Excerpts from a commonly used machine learning textbook, \textit{The Elements of Statistical Learning: Data Mining, Inference, and Prediction}, by Hastie, Tibshirani, and Friedman (2017). &
  Penalty functions, or regularization methods, express our prior belief that the type of functions we seek exhibit a certain type of smooth behavior, and indeed can usually be cast in a Bayesian framework. \\
Borges Dataset \cite{Adleme9463} &
  Contextual Confusion &
  Jorge Luis Borges is an Argentinian author known for blurring the lines between reality and fiction, often creating a sense of uncertainty and infinite possibilities. This dataset contains a collection of poems. Translated from Spanish to English. &
  I have vaguely looked for them in this old white, rectangular house, in the freshness of its two galleries, in the growing shadow cast by the pillars, in the timeless cry of the bird. \\
Baidu Tieba Ruozhiba Dataset \cite{LooksJuicy64} &
  Contextual Confusion &
  Ruozhiba is a popular online subforum within the Chinese community platform Baidu Tieba known for its linguistic intricacies and obscure jokes meant to entertain readers. Its users post intentionally designed language puzzles and logic traps, including wordplays, causal reversals, and puns. Translated from Mandarin Chinese to English. \cite{bai2024coig} &
  I was only born once. Why do I have to celebrate my birthday every year? \\
\hline
\end{tabularx}
\end{table}

\subsection{Equipment}

We recorded EEG data using a 64-channel eego™ system (ANT Neuro) at 512 Hz, controlled via the dedicated eego software \cite{eego64}. Binocular eye-tracking data was collected with Pupil Core glasses and the Pupil Capture software \cite{kassner2014pupil}. The experimental interface, including stimulus presentation and event marking, was implemented in PsychoPy \cite{peirce2019psychopy2}, augmented with custom scripts to log the precise on-screen coordinates of each displayed word for alignment with gaze data. To synchronize all streams from EEG, eye-tracking, and PsychoPy event markers, we employed the Lab Streaming Layer (LSL) framework \cite{kothe2024lab}, ensuring precise temporal alignment across modalities.

\newpage
\subsection{Participants}
\label{sec:participants}

We recruited 16 participants (10 female, 5 male, 1 non-binary; median age = 23.0 years, SD = 9.6). Of these, 11 (8 female, 3 male; median age = 28.0 years, SD = 11.3) completed the experiment with full EEG and eye-tracking datasets and were included in the final analysis. All participants had normal or corrected-to-normal vision and reported no neurological conditions. They were either native English speakers or demonstrated full professional proficiency. Each participant received a \$50 Amazon gift card as compensation. The study was approved by the Massachusetts Institute of Technology Committee on the Use of Humans as Experimental Subjects (protocol ID 21070000428).


\subsection{Experimental Protocol}

The experiment was conducted in a quiet, darkened laboratory room with no windows to minimize distractions and ensure high-quality EEG and eye-tracking recordings. Participants turned off all electronic devices, provided informed consent, and completed a background survey. A 64-channel EEG headset and Pupil Labs eye-tracking glasses were then fitted. Conductive gel was applied to each EEG electrode, and impedance levels were verified in the eego™ software to remain below 30 k$\Omega$. The eye tracker was calibrated using Pupil Labs software \cite{kassner2014pupil}. Before starting the main task, participants completed a training session of five paragraphs to familiarize themselves with the interface and procedure.

During the experiment, participants read short paragraphs presented sequentially on a computer screen with white text on a dark background (Figure \ref{fig:fig1}, left). Each paragraph remained visible for 10 seconds, allowing for complete reading before automatically advancing to the next (Figure \ref{fig:fig1}, right). A one-minute break was provided midway, between the 150th and 151st trials. The total experiment time, excluding setup, was approximately 60 minutes.

To incentivize participants to remain attentive, 60 of the 300 paragraphs (20\%) were followed by randomly distributed control questions. Each question was tailored to its respective paragraph and offered two response options, (a) and (b), which participants selected using the corresponding keyboard keys. The questions were designed to be easy when participants were focused. For instance, at trial 42, in the \textit{Factual Confusion} condition, the displayed paragraph read: "Oaks are strong, long-lived trees that are valued for their wood and ecological importance. They provide support for understanding and writing computer code in different programming languages." After 10 seconds, participants encountered the question: “What was the last paragraph talking about?” with answer choices (a) Oaks and (b) Cars, where the correct response was (a). These attention checks were used solely to ensure that participants were paying attention to the task and were not included in any further analysis.


\section{Analysis}

\begin{figure}[ht]
    \centering
    \includegraphics[width=\linewidth]{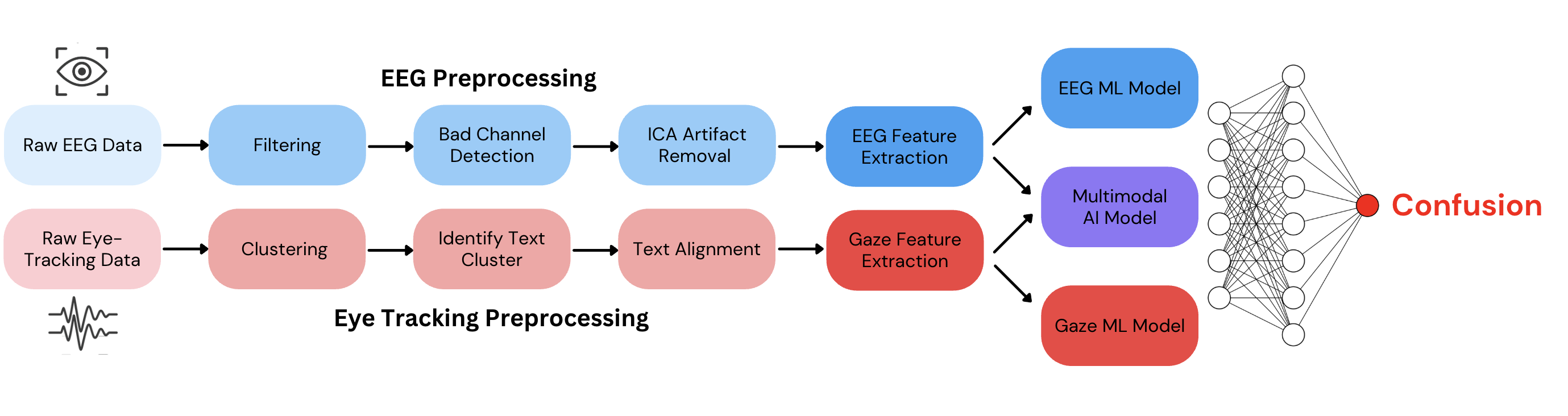}
    \caption{\textit{Overview of Data Analysis Pipeline.} Following data collection, we applied separate preprocessing pipelines for EEG and eye tracking data. EEG preprocessing included filtering, removal of bad channels, Independent Component Analysis (ICA) for artifact correction, and standardization. For eye tracking, preprocessing involved clustering raw gaze points into fixations, normalizing gaze coordinates, and aligning gaze events with the corresponding text regions. After preprocessing, we extracted modality-specific features and trained separate machine learning models for EEG and eye tracking. Finally, we fused both feature sets to build a multimodal model that integrated neural and gaze information.
}
    \label{fig:fig2}
\end{figure}

\subsection{EEG Preprocessing}

Our EEG preprocessing pipeline (Figure \ref{fig:fig2}) was implemented in Python using functions from SciPy and the MNE package \cite{2020SciPy-NMeth, gramfort2013meg}, and followed the guidelines of the PREP EEG preprocessing pipeline \cite{bigdely2015prep}. As part of the filtering process, we applied a finite impulse response (FIR) 60 Hz notch filter to remove power-line noise, a 1 Hz high-pass FIR filter to correct low-frequency drifts, and a 100 Hz low-pass FIR filter to suppress high-frequency noise.

To identify and remove bad channels, we first excluded those pre-flagged as problematic during data collection. We then applied a modified version of the PREP correlation criterion \cite{bigdely2015prep} to detect cross-conducting or noisy electrodes: pairwise correlations were calculated between each channel and all others; values outside the range 0.2 < $r$ < 0.99 were considered abnormal. Channels showing more than 80 \% abnormal correlations were excluded from further analysis for that participant.

To remove additional EEG artifacts, such as those associated with eye blinks, we performed Independent Component Analysis (ICA). Following the guidelines by Iriarte et al. \cite{iriarte2003independent}, we decomposed the EEG into 20 independent components, a compromise between capturing >80\% of explained variance and maintaining the effective rank of the data (i.e. the number of good channels). Components were visually inspected to remove artifacts related to ocular, muscular, and other non-EEG activity. The remaining independent components were then used to reconstruct the signals for subsequent analyses.

Finally, we segmented the data into 300 ten-second epochs using event markers recorded by PsychoPy. Each epoch was labeled as \textit{Control}, \textit{Factual Confusion}, or \textit{Contextual Confusion} (Table \ref{tab:table-1}). For consistency across channels, each was z-scored by subtracting its mean and dividing by its standard deviation (Figure \ref{fig:fig2}).

\subsection{EEG Classification}

The standardized EEG data from the retained channels was further segmented into overlapping 2-second windows, with a 50\% overlap between consecutive windows. This approach facilitated data augmentation (50\% increase in data) and addressed the limited dataset size by increasing the number of training samples. The resulting windowed dataset was structured with a shape of (trials × windows × 1024), where each window represented 2 seconds sampled at 512 Hz.

Next, we applied a series of feature extraction steps to analyze confusion-related patterns. Spectral powers were calculated for five standard frequency bands: delta (0.5–4 Hz), theta (4–8 Hz), alpha (8–13 Hz), beta (13–30 Hz), and gamma (>30 Hz). To improve frequency resolution and feature extraction, these bands were further subdivided into smaller sub-bands, with extracted powers included as features. Additionally, time-domain features (mean, standard deviation, kurtosis, skewness) and frequency-domain features (mean frequency, peak frequency) were extracted. This process produced a comprehensive feature set comprising 16 features for each of the four windows within each trial. 

Despite each paragraph in the reading task being similar in length, participants occasionally finished reading early, resulting in a lack of reading behavior near the end of the 10-second trial. To ensure that the EEG data reflected only the active reading period, we analyzed only the first 6 seconds of each trial. This exclusion mitigated potential post-activity noise resulting from early task completion. The resulting feature set was then split trial-wise into an 80/20 training and testing set, which was used as input for an XGBoost classifier \cite{chen2016xgboost}, configured with 100 estimators and a maximum depth of 60. For CNN models, we used preprocessed EEG data directly as input.

\subsection{Eye Tracking Preprocessing}

To improve the signal-to-noise ratio in our eye tracking data, we employed the DBSCAN algorithm \cite{ester1996density} to identify spatially coherent groups of gaze points and filter out isolated or noisy samples. After clustering, we selected the least noisy gaze cluster for each trial for downstream analysis, based on a custom scoring function that favors large, spatially coherent clusters with stable vertical gaze trajectories (defined in Eq.~\ref{eq:eq1}). To account for inter-individual variability in eye tracking behavior, we tuned the DBSCAN hyperparameters and the cluster selection procedure on a per-participant basis.

Formally, given raw gaze data $\mathcal{G} = \{(x_i, y_i, c_i)\}_{i=1}^N$, where each sample has an associated confidence score $c_i \geq \mathit{confidence\_threshold}$, we applied DBSCAN to obtain a set of clusters $\mathcal{C} = \{C_1, C_2, \dots, C_K\}$, excluding outliers. Each cluster $C_k = \{ (x_{k,1}, y_{k,1}), \dots, (x_{k,n_k}, y_{k,n_k}) \}$ was evaluated using a weighted scoring function combining its size $n_k$ and vertical uniformity score $u_k$:

\begin{equation}
\label{eq:eq1}
    C^* = \arg\max_{C_k \in \mathcal{C}} \; n_k \cdot w_{\text{size}} + u_k \cdot w_{\text{uniformity}}
\end{equation}

Here, the vertical uniformity score $u_k$ quantifies the vertical gaze stability within each cluster. Specifically, we divide the cluster into two temporal halves and compute the absolute difference between their mean $y$-coordinates:

\begin{equation}
\label{eq:eq2}
u_k = \left| \frac{1}{\lfloor n_k/2 \rfloor} \sum_{j=1}^{\lfloor n_k/2 \rfloor} y_{k,j} 
- \frac{1}{n_k - \lfloor n_k/2 \rfloor} \sum_{j=\lfloor n_k/2 \rfloor + 1}^{n_k} y_{k,j} \right|
\end{equation}

Lower values of $u_k$ indicate greater vertical consistency, implying less drift and potentially higher data quality. By optimizing Equation~\ref{eq:eq1}, we selected the cluster that maximized both size and vertical stability, thereby improving the robustness of subsequent gaze-based analyses.

\subsection{Eye Tracking Classification}

Similar to EEG classification pipeline, we then transformed the preprocessed gaze data into features including gaze velocity, fixation metrics, and gaze entropy \cite{shiferaw2019review}. These features were used to construct a tabular dataset for XGBoost model input. Additionally, to explore the deep learning approach, we applied a CNN directly to preprocessed eye tracking gaze coordinates, inspired by \cite{yin2018classification}. The details of the CNN implementation can be found in Appendix \ref{app:eye-tracking-classification}.

\subsection{EEG and Eye Tracking Integration (Ensemble Method)}

To effectively leverage the complementary insights offered by EEG and eye tracking data, we developed a multimodal integration strategy based on ensembling. The approach combined features derived from EEG used in XGBoost classification, and raw coordinate-based CNN inputs from eye tracking data.

In our fusion process, we used a weighted ensemble of the trained models, assigning an 80\% weight to the EEG model and 20\% to the eye tracking model. This weighting strategy was carefully designed to reflect the relative contributions of each modality in distinguishing confusion states, as informed by the classification performance metrics of the EEG-only (\ref{sec:results-eeg-classification}) and eye tracking-only models (\ref{sec:results-eye-tracking-classification}). The resulting multimodal model allows for the simultaneous incorporation of features from both modalities, creating a comprehensive framework for enhanced classification.


\section{Results}

In this section, we first visualize the N400 component and assess its statistical significance. We then align eye tracking data with the presented text to reconstruct participants’ gaze patterns. Finally, we present the results of training machine learning models for confusion detection.

\subsection{EEG: N400}

\begin{figure}[ht]
    \centering
    \includegraphics[width=16cm]{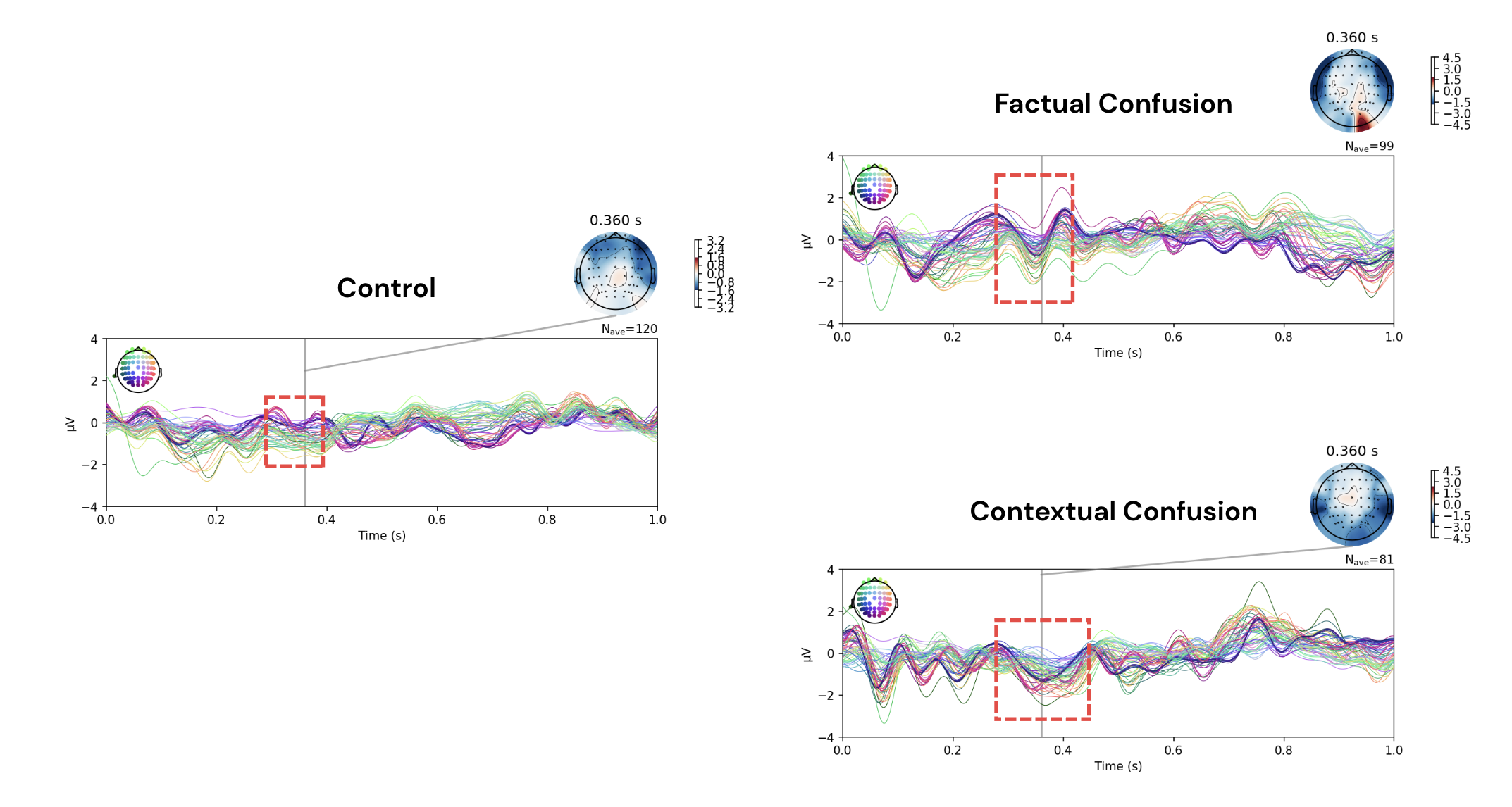}
    \caption{Trial-averaged EEG signals from 0 to 1 s post-stimulus onset for each condition, low-pass filtered at 15 Hz to emphasize event-related potentials (ERPs) in Participant P01. Red rectangles mark the time window around 360 ms, corresponding to the expected N400 component. At this latency, condition-specific differences are observed in both waveform morphology and scalp topography.
    Topographic maps at 360 ms are shown in the top-right corner of each panel. In the \textit{Control} condition, no prominent N400 response is visible; a weak negative deflection is observed primarily over the right temporal region. In the \textit{Factual Confusion} condition, a pronounced negative-going ERP is observed around 360 ms, with strong bilateral temporal negativity. The \textit{Contextual Confusion} condition also shows a negative deflection in the same time window, but with a more diffuse and attenuated spatial pattern, potentially reflecting temporal variability in confusion onset. Together, these results suggest condition-specific modulation of the N400 component, most robustly in response to factual incongruities.
    }
    \label{fig:fig3}
\end{figure}

In Figure~\ref{fig:fig3}, we present the trial-averaged EEG waveforms for each condition for one sample participant. In the \textit{Factual Confusion} condition, a prominent negative-going deflection emerges around 400 ms post-stimulus, consistent with the temporal and spatial characteristics of the N400 component \cite{vsovskic2022better,kutas2011thirty}. A comparable, though attenuated, negativity is observed in the \textit{Contextual Confusion} condition, with greater inter-trial variability suggesting less consistent onset timing. Notably, both confusion conditions exhibit enhanced negative activity over the left temporal region, a hallmark of the canonical N400 topography \cite{van2006neural,frishkoff2004frontal} (see Figure~\ref{fig:fig3}).

\begin{figure}[H]
    \centering
    \begin{subfigure}[t]{0.48\textwidth}
        \centering
        \includegraphics[height=6.0cm]{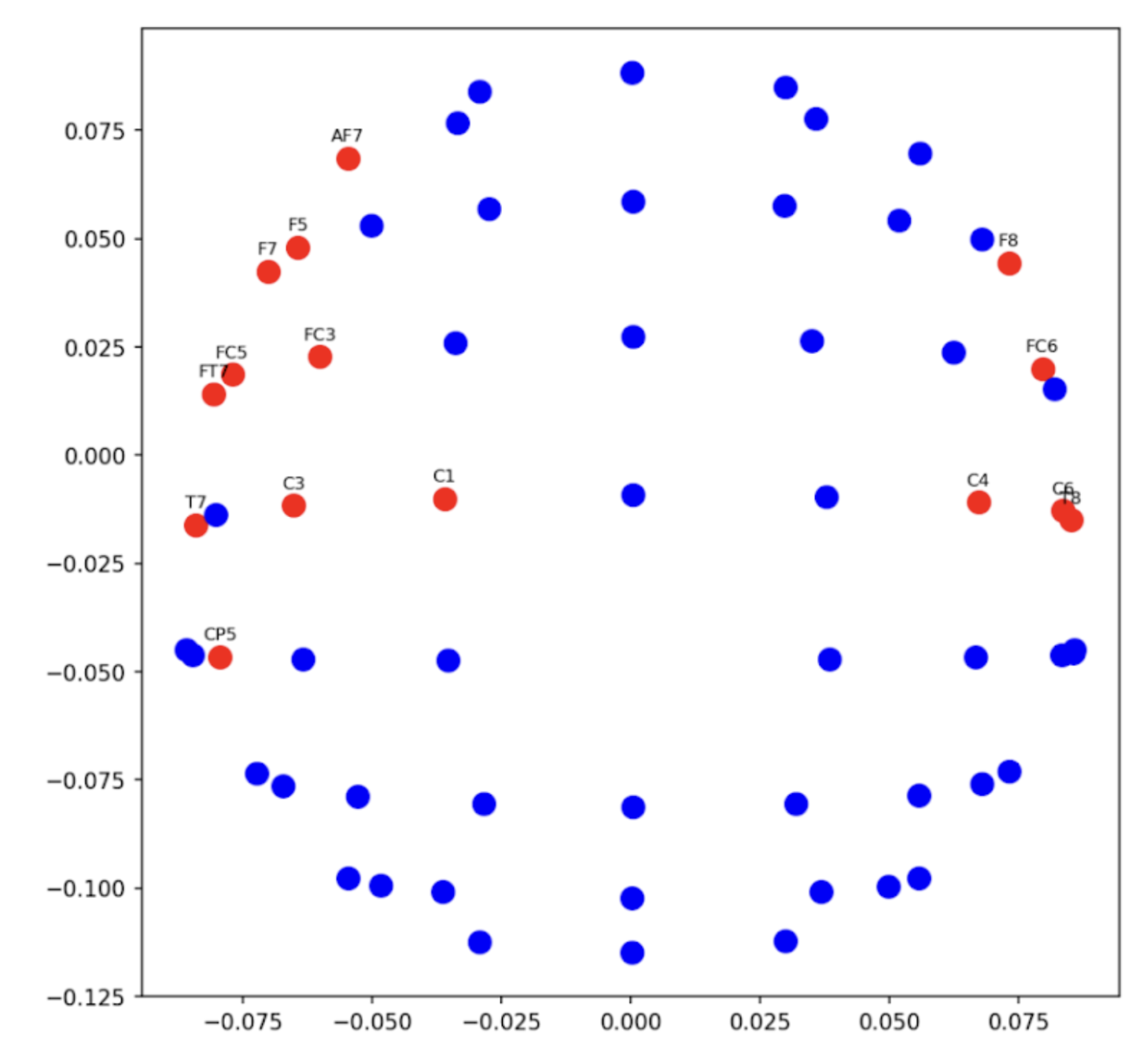}
        \label{fig:fig4a}
    \end{subfigure}
    \hfill
    \begin{subfigure}[t]{0.48\textwidth}
        \centering
        \includegraphics[height=6.2cm]{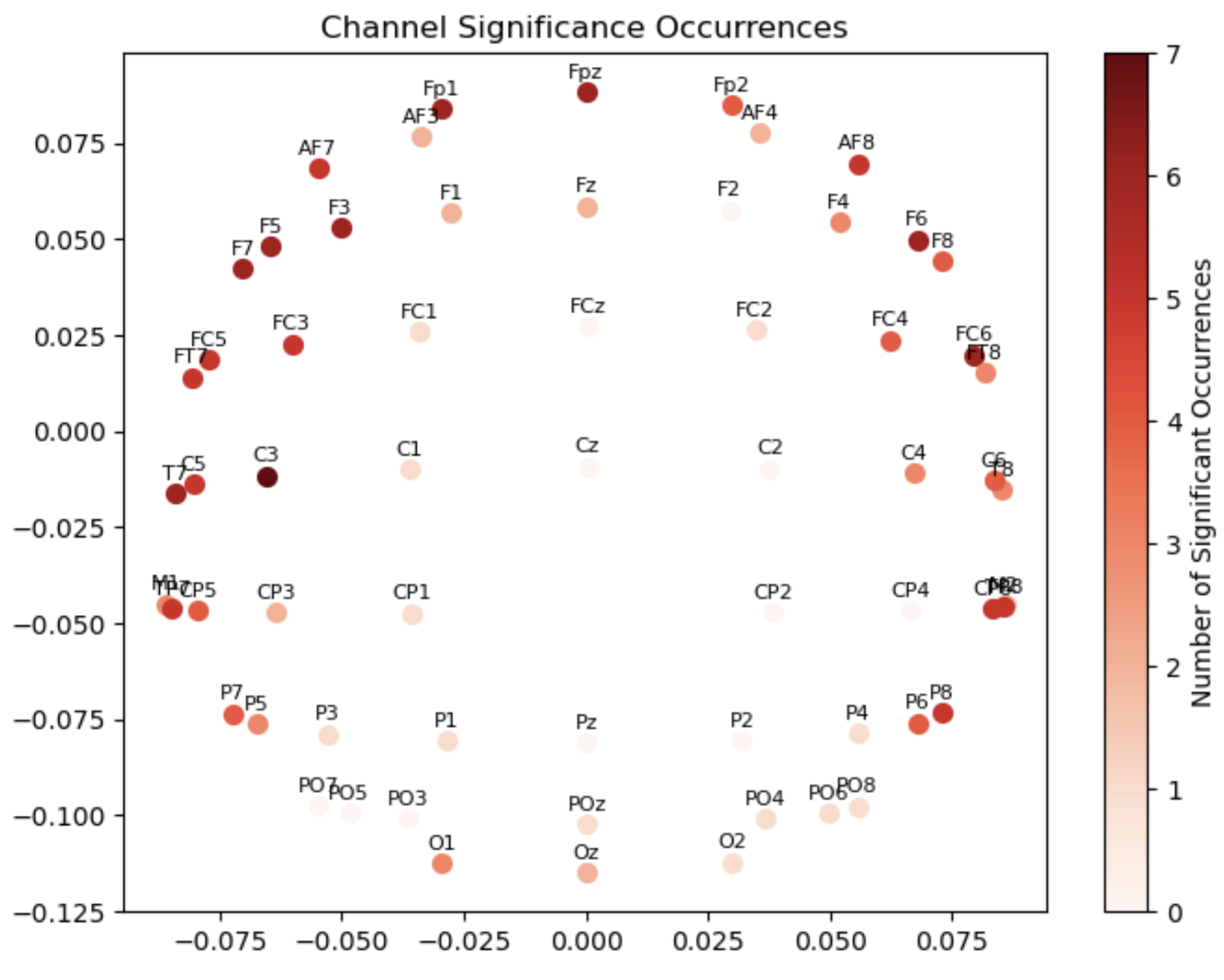}
        \label{fig:fig4b}
    \end{subfigure}
    \caption{\textit{Left}: Top-down view of EEG channel locations in Participant P01, with channels showing statistically significant band power differences (p < 0.02) highlighted in red. Most significant channels are concentrated in the temporal region. \textit{Right}: Across all 11 participants, the heatmap shows the number of times each channel exhibited statistically significant band power differences. Channels in the frontal-temporal regions are consistently the most influential ones.}

    \label{fig:fig4}
\end{figure}


Focusing on the time window surrounding the N400 component (350–450 ms), we computed the topological power spectrum across three main frequency bands, Alpha (8–12 Hz), Beta (12–30 Hz), and Gamma (30–45 Hz). One-way ANOVA tests were conducted on the spectral features across all channels, and the statistically significant ones were recorded.

We found that 9 out of 11 participants exhibited statistically significant N400 responses (p < 0.02) in the \textit{Factual Confusion} condition but not in the \textit{Control} condition. Similarly, 7 out of 11 participants showed a statistically significant N400 (p < 0.02) in the \textit{Contextual Confusion} condition. For full participant-level statistics, see Appendix~\ref{app:n400-significance}. Notably, the most discriminative features were localized to frontal, fronto-central, and temporal regions across all three frequency bands (Figure~\ref{fig:fig4}).

\begin{figure}[ht]
    \centering
    \includegraphics[width=\linewidth]{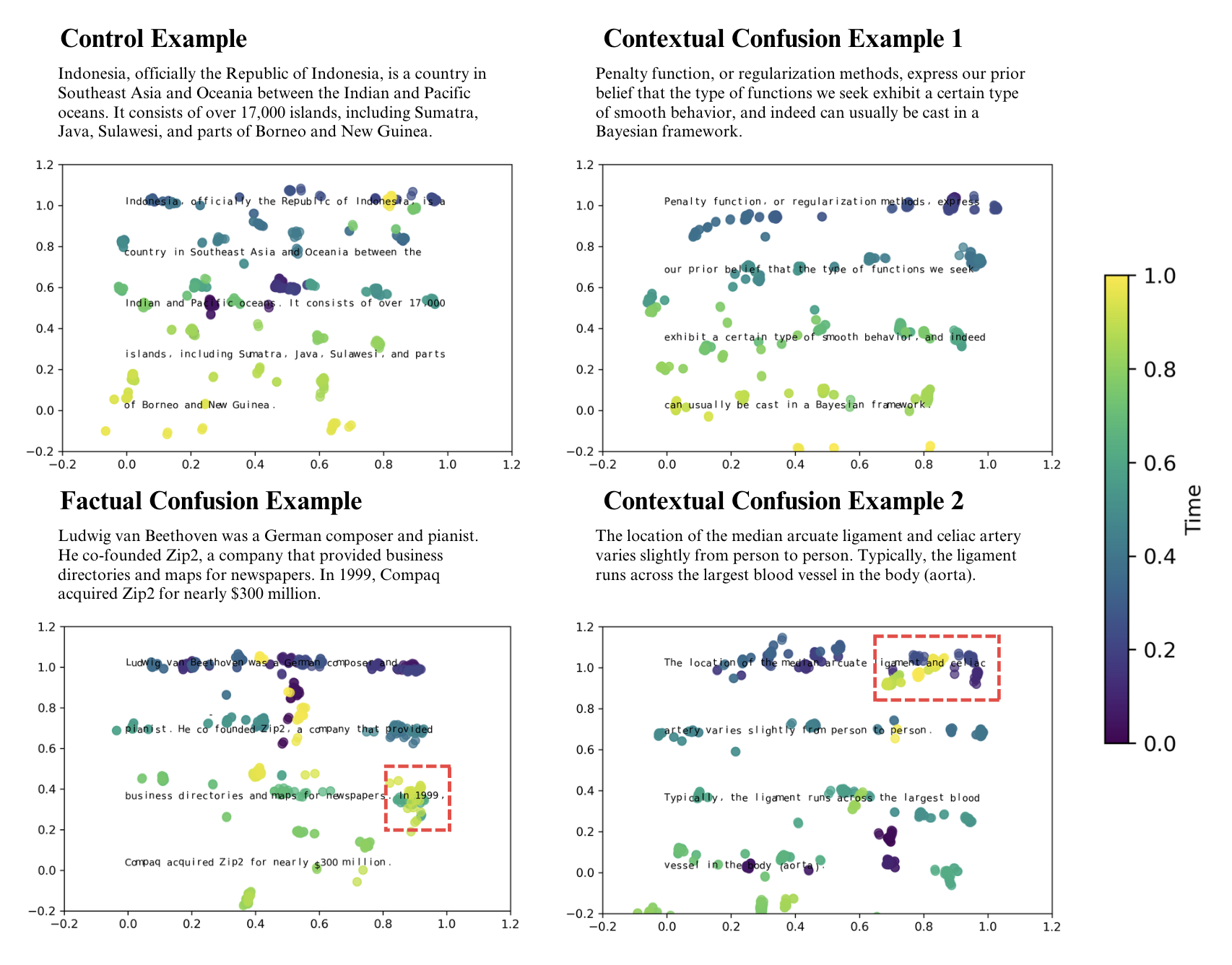}
    \caption{\textit{Examples of Eye Tracking and Text Alignment from Different Conditions.}
    \textit{Top left vs. Bottom left}: In the \textit{Factual Confusion} trial (bottom left), dense gaze points (highlighted in red rectangles) are evident, in contrast to the control trial (top left), which exhibits a more uniform gaze point distribution. We can hypothesize that the participant experienced confusion regarding the mention of "Zip2" and the date "1999," which conflicted with the name "Beethoven."
    \textit{Top right vs. Bottom right}: Both paragraphs aim to evoke \textit{Contextual Confusion}. However, example trial 1 (top right) results in fewer clustered gaze points and a more even distribution compared to example trial 2 (bottom right). This difference may be attributed to the participant's prior knowledge of machine learning (top right), which was not the case for medical science (bottom right), as confirmed by the pre-experiment survey.
    }
    \label{fig:fig5}
\end{figure}

\subsection{Eye Tracking Alignment}

In Figure~\ref{fig:fig5}, we present examples of the gaze points aligned with the text displayed on the screen. This alignment process mirrors the methodology used in the ZuCo dataset \cite{hollenstein2018zuco, hollenstein2019zuco}, ensuring consistency and accuracy in linking gaze data with textual information.



\newpage
\subsection{Classification Results}

\begin{figure}[ht]
    \centering
    \includegraphics[width=12cm]{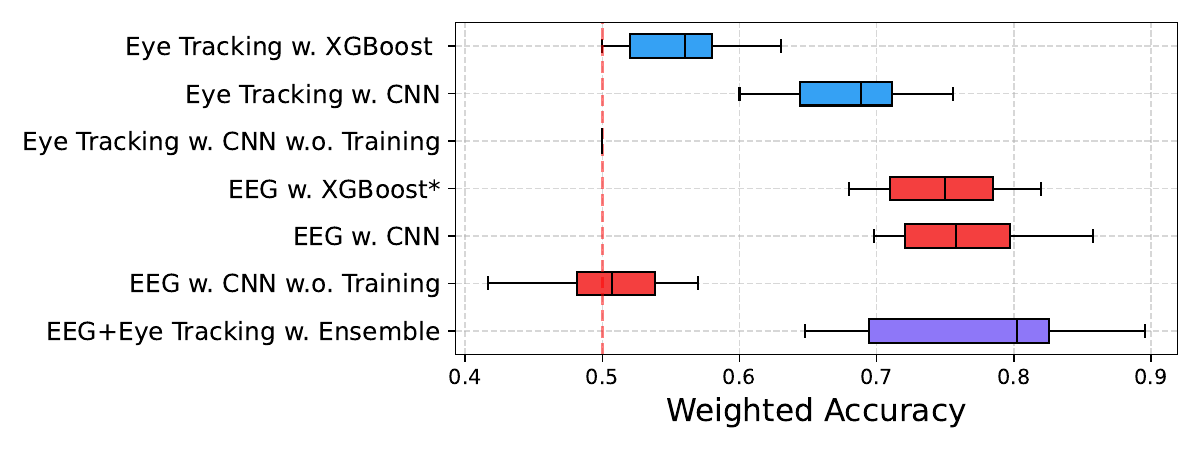}
    \caption*{\small\textit{* Trained only on features from temporal region electrodes}}
    \caption{Box-and-whisker plots showing the weighted testing accuracy of the evaluated machine learning models. The horizontal black line inside each box marks the \textbf{median} accuracy, while the box edges represent the interquartile range. Whiskers extend to the minimum and maximum observed values, excluding outliers. The red dashed line indicates chance-level performance.
    }
    \label{fig:fig7}
\end{figure}

We summarize the classification performance of EEG-, eye-tracking-, and multimodal models using XGBoost and CNN   in Table~\ref{tab:table-3} and Figure~\ref{fig:fig7}. For each model, we report the mean balanced training accuracy, mean balanced testing accuracy, and best balanced testing accuracy across participants. EEG-based models consistently outperformed eye-tracking-only models, and multimodal models also demonstrated significantly higher performance than eye-tracking alone. While the multimodal models did not significantly outperform EEG-only models, they still achieved a mean improvement of 3.45\% in testing accuracy and a median improvement of 4.41\%. Participant-level classification performance is reported in Appendix ~\ref{app:ensembling-accuracy}. 
Since untrained CNNs can function similarly to random feature extractors \cite{saxe2011random}, we investigated their ability to extract patterns from data. Our findings revealed that the accuracy of the untrained CNN was near chance levels, reinforcing the conclusion that training on the dataset is essential for effectively identifying confusion and discerning meaningful patterns.

\begin{table}[ht]
\caption{Weighted Classification Accuracies Summary for Temporal Region Electrodes}
\label{tab:table-3}
\begin{tabularx}{\textwidth}{Xccc}
\hline
\textbf{Modalities w. Model} & \textbf{Mean Training} & \textbf{Mean Testing} & \textbf{Best Testing} \\
\hline
\textbf{Eye Tracking w. XGBoost}             & 89.63\% & 55.00\% & 67.00\% \\
\textbf{Eye Tracking w. CNN}                 & 90.00\% & 67.34\% & 75.56\% \\
\textbf{Eye Tracking w. CNN w.o. Training}   & -       & 50.00\% & 50.00\% \\
\textbf{EEG w. XGBoost*}                     & 82.64\% & 73.84\% & 81.66\% \\
\textbf{EEG w. CNN}                          & 86.36\% & 72.93\% & 85.78\% \\
\textbf{EEG w. CNN w.o. Training}            & -       & 50.10\% & 56.94\% \\
\textbf{EEG+Eye Tracking w. Ensembling}      & 81.72\% & 77.29\% & 89.55\% \\
\hline
\end{tabularx}
\end{table}

\subsubsection{EEG Classification}
\label{sec:results-eeg-classification}

For the EEG-based models, both XGBoost and CNN classifiers demonstrated strong performance in decoding confusion states. The XGBoost model, trained on a subset of channels in the frontal-temporal area, achieved an average testing accuracy of 73.84\% and the best testing accuracy of 81.66\%. The CNN model showed slightly lower average testing performance at 72.93\% but outperformed XGBoost in terms of the best testing accuracy, reaching 85.78\%. These results highlight neural patterns associated with confusion can be well captured by EEG signals.

\subsubsection{Eye Tracking Classification}
\label{sec:results-eye-tracking-classification}

Eye-tracking-based models revealed substantial performance differences between traditional and deep learning approaches. The XGBoost model, which used engineered features such as gaze velocity and fixation metrics, achieved an average test accuracy of 55.00\% and a best-case accuracy of 67.00\%, only modestly above the 50.00\% chance level for a balanced binary classification task. In contrast, the CNN model, trained directly on raw eye-tracking coordinates, significantly outperformed the feature-based approach, reaching an average test accuracy of 67.34\% and a best test accuracy of 75.56\%.

These results suggest that raw eye tracking data encapsulates richer spatial and temporal information, which deep learning methods can leverage more effectively than traditional approaches. Furthermore, the data appears to provide valuable insights into the reader’s confusion state, as evidenced by clusters of fixations on specific words that may indicate difficulties in semantic processing.

\subsubsection{EEG and Eye Tracking Integration (Ensemble Method)}

Finally, the ensemble model, which integrates EEG and eye tracking data using an 80-20 weighted fusion strategy, yielded the best overall performance. The multimodal approach achieved an average testing accuracy of 77.29\% and the best testing accuracy of 89.55\%, outperforming EEG-only and eye-tracking-only models. This improvement highlights the complementary nature of EEG and eye tracking data, where temporal dynamics of EEG and the spatial gaze patterns from eye tracking collectively enhance the model’s ability to distinguish between confusion states. 

\section{Discussion}

\subsection{Key Findings and Their Significance}
This study is, to our knowledge, the first to investigate confusion arising during the natural reading flow of paragraph-length text, capturing the integrative and context-dependent processes absent from traditional word-by-word paradigms \cite{delogu2017teasing,van2006neural,frishkoff2004frontal,kutas1980reading}. We show that confusion can be reliably elicited and detected in ecologically valid reading scenarios, as evidenced by robust neural and behavioral markers.

To operationalize confusion, we introduced two systematically defined subtypes: \textit{Factual Confusion}, triggered by violations of widely known facts, and \textit{Contextual Confusion}, stemming from insufficient background knowledge. Both elicited well-established signatures of semantic processing difficulty, including N400 ERPs and distinctive gaze fixation patterns. \textit{Factual Confusion} proved easier to detect, producing more consistent N400 responses and spatially coherent gaze clusters, whereas \textit{Contextual Confusion} exhibited greater inter-participant variability, likely reflecting individual differences in prior knowledge. Notably, models trained on both subtypes generalized well across them, suggesting that shared cognitive processes underlie different forms of confusion.

Our multimodal machine learning pipeline, integrating EEG and eye tracking, achieved 77.29\% mean testing accuracy and 89.55\% best-subject accuracy, outperforming prior EEG-only approaches with similar non-interactive stimuli (56–69\% \cite{wang2013using,kopparapu2023spatial}) and matching results from interactive tasks that benefit from explicit behavioral labels \cite{xu2023confused,zhou2018confusion,salminen2019confusion,sims2020neural}. Crucially, no ground-truth user responses were required, demonstrating feasibility for passive information-intake contexts such as education, adaptive learning, and assistive reading technologies.

Eye tracking contributed complementary spatial markers, e.g., fixation clustering on semantically challenging words, that paralleled EEG temporal patterns. Multimodal fusion improved classification accuracy beyond either modality alone, consistent with evidence that cognitive states are distributed across neural, behavioral, and physiological channels \cite{taher2015multimodal}. This highlights opportunities for end-to-end multimodal representation learning to capture subtler or more extended confusion dynamics.

Finally, our results reinforce the N400 component’s relevance as a biomarker for confusion in natural reading. Significant N400 responses were observed in most participants (9/11 for \textit{Factual}, 7/11 for \textit{Contextual}), with frontal-temporal regions most discriminative for classification. Although paragraph-level presentation precludes precise single-word alignment, the elicited dynamics parallel those from linguistic incongruence studies \cite{kutas2011thirty,he2024multivariate,rabs2022situational,lau2016direct,white2009wait,van2006neural}, underscoring the applicability of N400 in ecologically valid contexts.

In sum, we establish that confusion during naturalistic paragraph reading can be robustly detected using a multimodal EEG–eye tracking framework. By combining high-resolution neural and behavioral signals, we advance the state of the art beyond prior work constrained by artificial stimuli, controlled tasks, or unimodal input.

\subsection{Implications for Adaptive Brain-Computer Interfaces}

Our findings reveal important opportunities for translating confusion detection into practical applications that enhance human learning, accessibility, and human-computer interaction. Confusion, long recognized as a driver for deeper cognitive processing, has been shown to significantly predict learning outcomes, as confused learners have shown up to 46\% improvement in post-instruction gains in intelligent tutoring systems (Cohen’s d = .64) \cite{craig2012confusion}. These studies treat confusion not as an undesirable state, but as a critical signal for adaptive support. Recognizing and responding to confusion in real time can make learners more aware of knowledge gaps, triggering reflection and facilitating more robust comprehension and retention \cite{craig2012confusion,sadras2023post}. Our work, by providing a confusion detection model during naturalistic reading, creates opportunities for integrating such mechanisms into next-generation educational platforms.

Our findings support the feasibility of incorporating confusion as a metric in closed-loop BCIs designed to adapt to the user's cognitive state, similar to NeuroChat, a neuroadaptive AI tutor that adapts its response style based on the user's cognitive engagement \cite{Baradari2025-sd}. We found that confusion detection is particularly effective when leveraging EEG signals from temporal areas - using only temporal electrodes improved test accuracy from 45.65\% to 73.84\%. This opens the possibility of deploying lower-density, wearable EEG systems focused on temporal areas for practical, everyday use in classrooms, online learning use cases, and accessibility contexts. Eye tracking integration further enhances system sensitivity, enabling seamless and unobtrusive monitoring that supports more dynamic and inclusive learning experiences.

Additionally, our experimental methodology enables the temporal alignment of eye gaze and textual content using wearable eye tracking devices, providing the foundation for future studies that could resolve word-level comprehension dynamics during natural reading. This capability creates opportunities for longitudinal studies and large-scale data collection, which might facilitate the development of foundation models that connect brain signals with language representations. For instance, the distinct N400 signatures we observed for different types of confusion (\textit{Control}, \textit{Factual Confusion}, \textit{Contextual Confusion}) indicate that neural responses to semantic or contextual novelty can be systematically characterized. Building on this, future research could directly link these neural responses to computational language model predictions, offering a framework to examine how unexpected words or conceptual incongruities are represented both in the human brain and in large language models. Such neuroadaptive interfaces could ultimately inform the development of LLMs that are not only contextually aware but also sensitive to users' comprehension states — enabling more personalized, empathetic, and effective human-AI interaction. These advances hold significant promise for closing the loop between user state, system feedback, and learning outcomes, fostering more responsive, inclusive, and cognitively aligned interfaces.

\subsection{Limitations and Future Directions}
While our study demonstrates the feasibility of multimodal confusion detection in natural reading flow, several limitations and important avenues for future research remain. First, the detection of \textit{Contextual Confusion} is fundamentally constrained by variability in individuals’ prior knowledge. Text samples that are confusing for one participant may be readily understood by another one, reflecting differences in domain expertise, background, or even transient attention states \cite{abdelaal2014relationship}. Our results indicate that, while \textit{Factual} and \textit{Contextual Confusion} both elicit robust neural and behavioral signatures (see Figure \ref{fig:fig3}, Figure \ref{fig:fig5}), \textit{Contextual Confusion} exhibits greater variability across individuals and domains. For example, subject-specific analysis across medical, machine learning, philosophy, and literature domains revealed that participants’ prior self-reported expertise influenced confusion detection. This underscores the importance of incorporating individualized knowledge profiles into future adaptive BCI systems - a direction that aligns with advances in personalized learning and neuroadaptive technologies \cite{kosmyna2019attentivu,sadras2023post,hollenstein2021decoding}. Ultimately, more granular or dynamic definitions of confusion will be necessary to capture the full spectrum of this cognitive state as it naturally occurs.

Another limitation of our study was the robustness of wearable eye tracking. While integrating eye tracking improved classification accuracy by 3.45\%, highlighting its value as a complementary modality, the quality of wearable eye tracking data posed unique challenges for us. Compared to table-mounted systems used in datasets like ZuCo \cite{hollenstein2018zuco,hollenstein2019zuco}, wearable versions of eye tracking are more susceptible to head movement, imperfect alignment, and individual differences in head shape or ocular physiology. These factors, as well as variable calibration quality, introduced noise and limited the reliability of fixation-based features. Despite these constraints, our results demonstrate the promise of combining wearable eye tracking with EEG for real-world cognitive assessment. To mitigate data quality issues, we implemented several measures: (1) we provided a flexible 3D-printed chin rest, though this proved uncomfortable during early testing stage and was removed, and (2) we performed manual, participant-specific tuning of preprocessing parameters, such as filtering and clustering, since standardized pipelines for wearable eye tracking remain underdeveloped \cite{geller2020gazer,ester1996density}. Future work will benefit from advances in both hardware development and automated, robust signal processing techniques for eye tracking in dynamic settings.

Finally, limited data scale remains a central challenge for machine learning-based confusion detection. Although data were collected from 11 participants reading 300 trials each, the dataset size was still constrained for training advanced machine learning models such as deep neural networks. Data augmentation through windowing and trial segmentation improved performance, but as recent large-scale EEG research shows, models benefit substantially from larger and more diverse datasets, sometimes spanning hundreds of hours of recordings \cite{sato2024scaling}. To fully realize the potential of multimodal, brain-aligned models for confusion detection, future research should prioritize the large-scale, longitudinal collection of neural and behavioral signals, enabling the development of robust, generalizable, and human-centered foundation models.

In summary, addressing these limitations by developing more individualized confusion taxonomies, improving wearable sensing and signal processing, and scaling up data collection will be critical for advancing the practical deployment of neuroadaptive learning systems and cognitive interfaces.

\section{Conclusion}

In this work, we present a multimodal framework for detecting confusion while participants are reading paragraphs in a high ecological validity setting, leveraging EEG and eye tracking devices. Our results demonstrate that confusion elicits distinct behavioral and neural markers, including fixation patterns and N400 event-related potentials, with temporal region electrodes providing the most informative neural features. By training and integrating XGBoost and CNN models on each modality, we achieved robust classification performance, reaching a best-subject accuracy of 89.55\% and an average accuracy of 77.29\%.
This approach highlights the value of combining neural and behavioral signals for cognitive state inference. We envision that such systems can enable next-generation accessibility tools and neuroadaptive interfaces in diverse educational and real-world settings. The development of large-scale, longitudinal datasets using wearable devices will further enhance the potential of the proposed framework, paving the way for deeper integration between human neural signals and language models and supporting the evolution of general-purpose augmentation tools for cognitive and educational applications.

\section{Acknowledgements}
We thank Ashley Gallitto for designing and printing our chin rest. We thank Emmie Fitz-Gibbon for helping with the experiment setup. 

\newpage
\bibliographystyle{unsrt}  
\bibliography{references}  

\newpage
\appendix


\newpage
\section{EEG-Paragraph Examples}
\label{app:paragraph-examples}

\begin{figure}[ht]
    \centering
    
    \begin{subfigure}[b]{\textwidth}
        \centering
        \includegraphics[height=5cm]{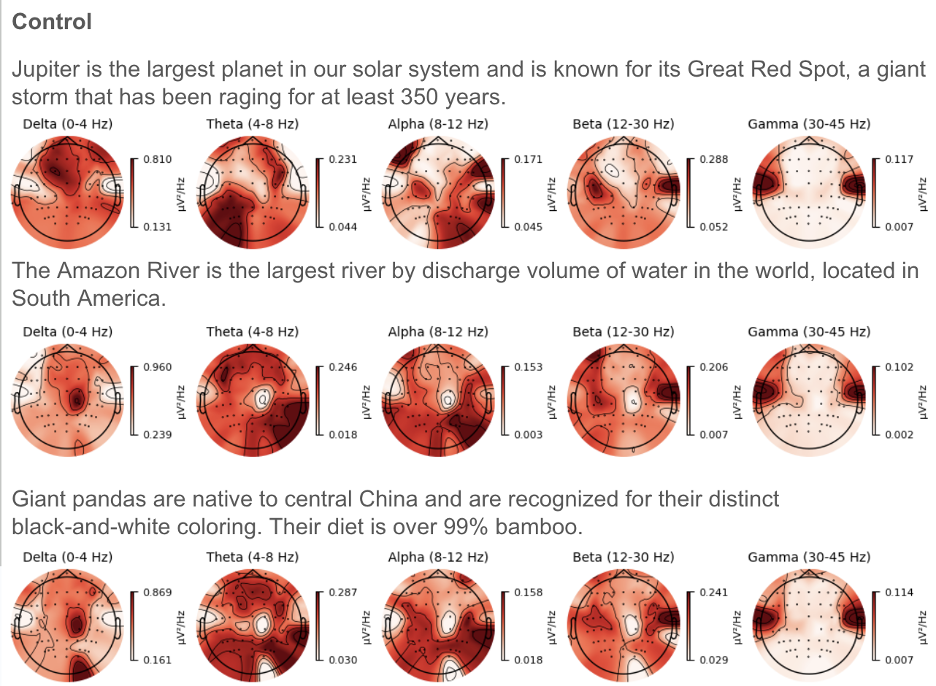}
        \caption{Control class}
        \label{fig:fig_b1}
    \end{subfigure}
    \hfill
    \begin{subfigure}[b]{\textwidth}
        \centering
        \includegraphics[height=5cm]{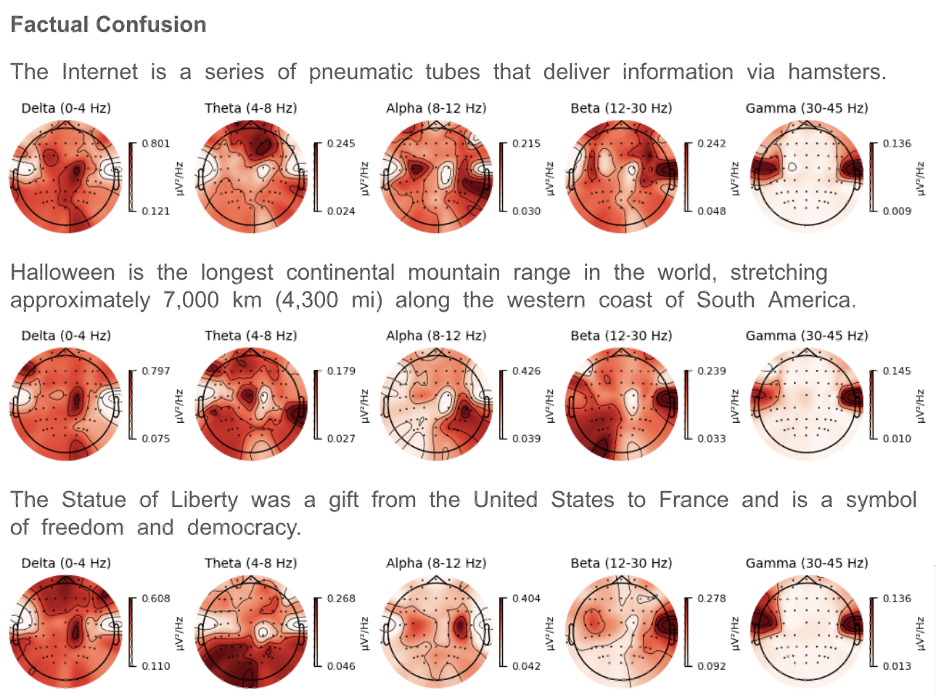}
        \caption{Factual Confusion class}
        \label{fig:fig_b2}
    \end{subfigure}
    \hfill
    \begin{subfigure}[b]{\textwidth}
        \centering
        \includegraphics[height=5cm]{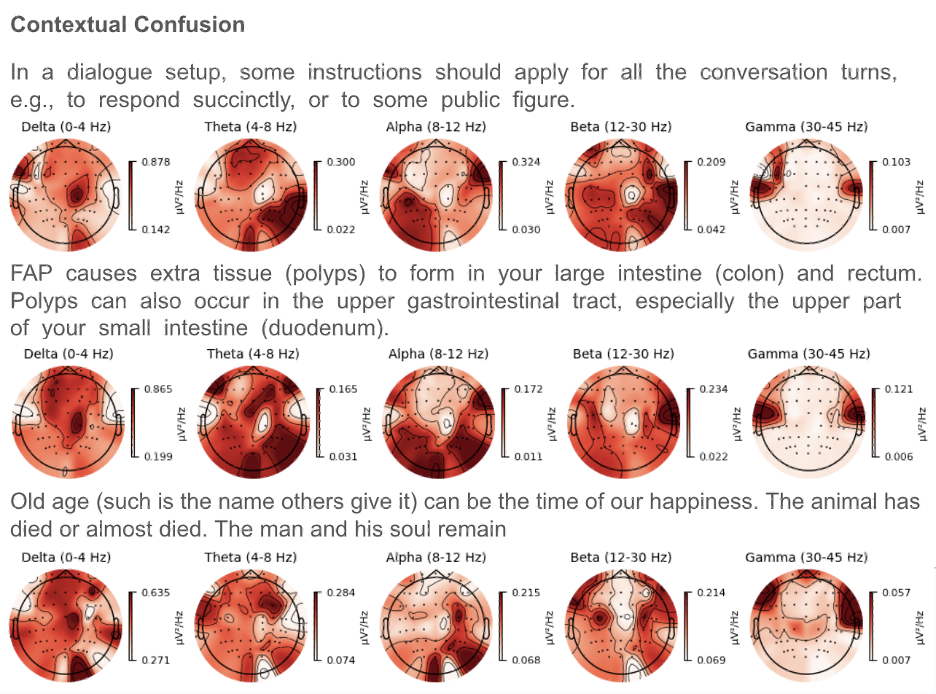}
        \caption{Contextual Confusion class}
        \label{fig:fig_b3}
    \end{subfigure}
    
    \caption{Topographic EEG maps of three example trials for each class: (a) Control, (b) Factual Confusion, (c) Contextual Confusion from one subject (p01). Each row corresponds to a single text trial, with corresponding paragraph content displayed above. For each trial, we illustrate the neural responses across five canonical EEG frequency bands: Delta (0–4 Hz), Theta (4–8 Hz), Alpha (8–13 Hz), Beta (13–30 Hz), and Gamma (30–45 Hz). Warmer colors indicate higher relative power over scalp regions. They demonstrate the variability in topographic distribution of EEG spectral power, supporting the presence of distinct neural signatures across different reading-induced confusion conditions.
    }
    \label{fig:combined_eeg_examples}
\end{figure}

\clearpage
\section{EEG N400 Significance}
\label{app:n400-significance}
\begin{table}[h]
\caption{p-values from N400 significance testing across 11 participants for two comparisons: \textit{Factual Confusion} vs. \textit{Control}, and \textit{Contextual Confusion} vs. \textit{Control}.}
\label{tab:my-table}
\begin{tabular}{l|rr}
\hline
 & \multicolumn{1}{l}{p-values, factual vs. control (all electrodes)} & \multicolumn{1}{l}{p-values, contextual vs. control (all electrodes)} \\
\hline
\textbf{P01} & \cellcolor[HTML]{B7E1CD}\textless{}0.02 & \cellcolor[HTML]{B7E1CD}\textless{}0.02 \\
\textbf{P02} & \cellcolor[HTML]{B7E1CD}\textless{}0.02 & \cellcolor[HTML]{B7E1CD}\textless{}0.02 \\
\textbf{P03} & \cellcolor[HTML]{B7E1CD}\textless{}0.02 & 5.43E-01                                \\
\textbf{P04} & \cellcolor[HTML]{B7E1CD}\textless{}0.02 & \cellcolor[HTML]{B7E1CD}\textless{}0.02 \\
\textbf{P05} & \cellcolor[HTML]{B7E1CD}\textless{}0.02 & \cellcolor[HTML]{B7E1CD}\textless{}0.02 \\
\textbf{P06} & \cellcolor[HTML]{B7E1CD}\textless{}0.02 & 2.82E-02                                \\
\textbf{P07} & \cellcolor[HTML]{B7E1CD}\textless{}0.02 & 1.12E-01                                \\
\textbf{P08} & 1.32E-01                                & \cellcolor[HTML]{B7E1CD}\textless{}0.02 \\
\textbf{P09} & \cellcolor[HTML]{B7E1CD}\textless{}0.02 & \cellcolor[HTML]{B7E1CD}\textless{}0.02 \\
\textbf{P10} & \cellcolor[HTML]{B7E1CD}\textless{}0.02 & \cellcolor[HTML]{B7E1CD}\textless{}0.02 \\
\textbf{P11} & \cellcolor[HTML]{B7E1CD}\textless{}0.02 & \cellcolor[HTML]{B7E1CD}\textless{}0.02 \\
\hline
\end{tabular}
\end{table}

\clearpage
\section{Eye Tracking Preprocessing}
\label{app:eye-tracking-preprocessing}
\begin{figure}[ht]
    \centering
    \includegraphics[height=10cm]{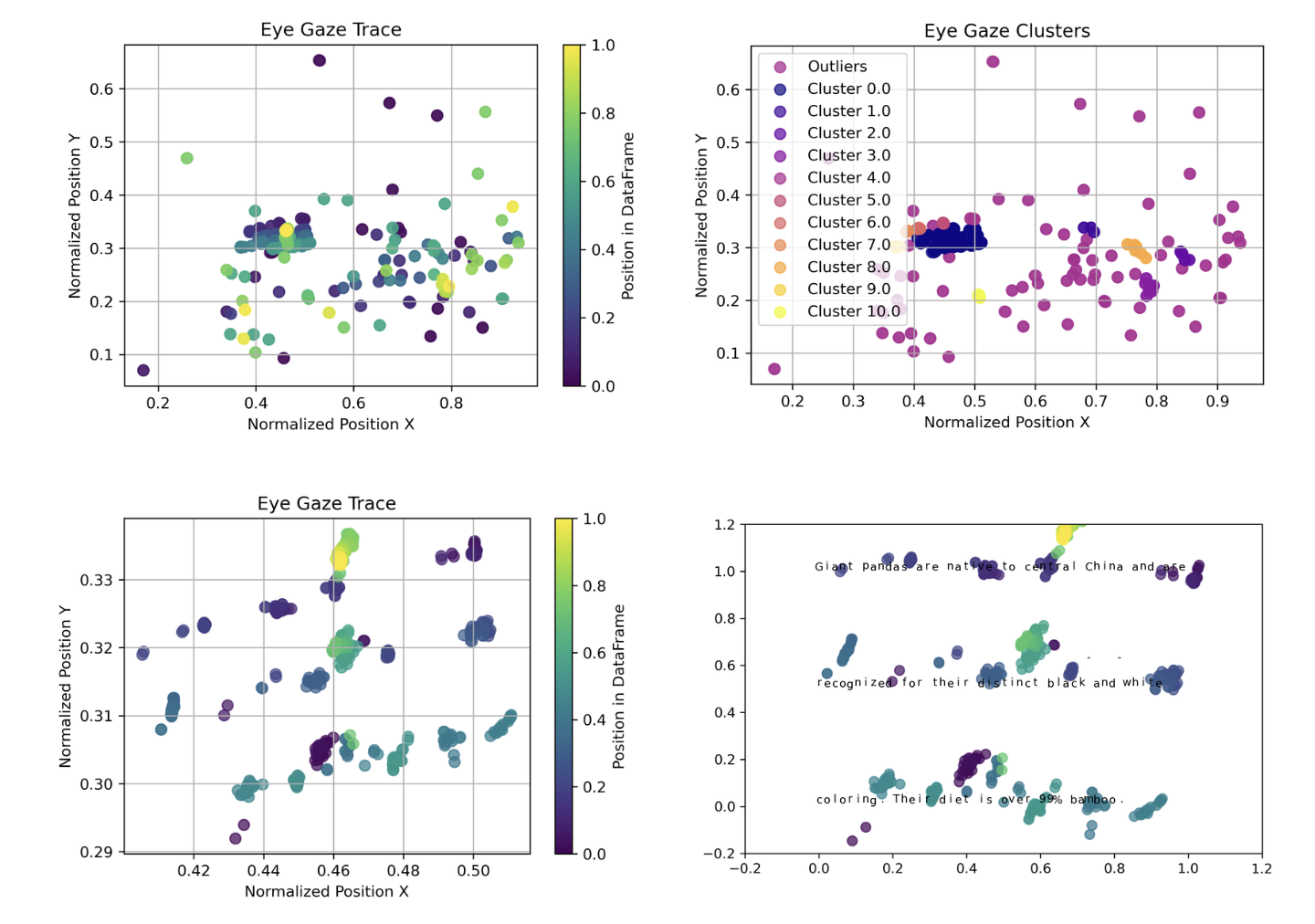}
    \caption{\textit{An Example Trial of Eye Tracking Preprocessing for Participant P01.} This is a control trial with text: "Chocolate is made from cocoa beans, which are the dried and fermented seeds of the cacao tree. It has been consumed as a drink for most of its history, especially in ancient Mesoamerica." \textit{Top Left:} Raw Gaze Coordinates of the Example Trial; \textit{Top right:} To find the gaze points related to text-reading and reject the noisy ones, we clustered gaze coordinates using DBSCAN [21] and calculated cluster scores according to equation \ref{eq:eq1}; \textit{Bottom Left:} Correct Cluster Identified; \textit{Bottom Right:} Alignment with Original Screen.
}
    \label{fig:fig_d1}
\end{figure}

\clearpage
\section{EEG Electrode p-values}
\label{app:eeg-p-values}

\begin{table}[ht]
\centering
\tiny
\caption{Each EEG channel's significance (p-value) as determined by comparing the channel’s average power bands intensity between different conditions for all participants}
\caption{}
\label{tab:tab_e1}
\begin{tabular}{lrrrrrrrrrrr}
\textbf{sbj} & \multicolumn{1}{l}{\textbf{p01}} & \multicolumn{1}{l}{\textbf{p02}} & \multicolumn{1}{l}{\textbf{p03}} & \multicolumn{1}{l}{\textbf{p04}} & \multicolumn{1}{l}{\textbf{p05}} & \multicolumn{1}{l}{\textbf{p06}} & \multicolumn{1}{l}{\textbf{p07}} & \multicolumn{1}{l}{\textbf{p08}} & \multicolumn{1}{l}{\textbf{p09}} & \multicolumn{1}{l}{\textbf{p10}} & \multicolumn{1}{l}{\textbf{p11}} \\
\textbf{Fp1} & \cellcolor[HTML]{B7E1CD}0.00001 & \cellcolor[HTML]{B7E1CD}0.00163 & \cellcolor[HTML]{B7E1CD}0.00023 & 0.46636 & \cellcolor[HTML]{B7E1CD}0.01249 & \cellcolor[HTML]{B7E1CD}0.00840 & 0.91681 & 0.07262 & 0.84630 & 0.24830 & \cellcolor[HTML]{B7E1CD}0.00002 \\
\textbf{Fpz} & \cellcolor[HTML]{B7E1CD}0.01939 & 0.34133 & \cellcolor[HTML]{B7E1CD}0.00043 & 0.39892 & \cellcolor[HTML]{B7E1CD}0.01640 & \cellcolor[HTML]{B7E1CD}0.01716 & 0.94411 & 0.10300 & 0.22479 & 0.31887 & \cellcolor[HTML]{B7E1CD}0.00011 \\
\textbf{Fp2} & \cellcolor[HTML]{B7E1CD}0.00020 & 0.24085 & 0.13418 & 0.05247 & \cellcolor[HTML]{B7E1CD}0.00553 & 0.12903 & 0.76240 & 0.13958 & 0.11282 & 0.31033 & \cellcolor[HTML]{B7E1CD}0.00020 \\
\textbf{F7} & \cellcolor[HTML]{B7E1CD}0.00002 & 0.13301 & \cellcolor[HTML]{B7E1CD}0.02169 & 0.26676 & 0.20112 & \cellcolor[HTML]{B7E1CD}0.00474 & 0.50845 & 0.29170 & 0.10165 & \cellcolor[HTML]{B7E1CD}0.00995 & \cellcolor[HTML]{B7E1CD}0.00086 \\
\textbf{F3} & \cellcolor[HTML]{B7E1CD}0.01546 & \cellcolor[HTML]{B7E1CD}0.01080 & \cellcolor[HTML]{B7E1CD}0.02164 & 0.12697 & 0.23819 & 0.03304 & \cellcolor[HTML]{B7E1CD}0.00129 & 0.52648 & 0.41985 & \cellcolor[HTML]{B7E1CD}0.02015 & \cellcolor[HTML]{B7E1CD}0.00150 \\
\textbf{Fz} & \cellcolor[HTML]{B7E1CD}0.01664 & 0.70112 & 0.40656 & 0.48006 & 0.82827 & 0.08830 & 0.26733 & 0.83714 & 0.52629 & 0.44104 & \cellcolor[HTML]{B7E1CD}0.00015 \\
\textbf{F4} & 0.04908 & \cellcolor[HTML]{B7E1CD}0.00591 & \cellcolor[HTML]{B7E1CD}0.01323 & 0.11888 & 0.52480 & 0.31125 & 0.63610 & 0.62488 & 0.03662 & 0.09634 & \cellcolor[HTML]{B7E1CD}0.01693 \\
\textbf{F8} & \cellcolor[HTML]{B7E1CD}0.00000 & \cellcolor[HTML]{B7E1CD}0.00270 & 0.66625 & \cellcolor[HTML]{B7E1CD}0.01586 & 0.05875 & 0.11564 & 0.58120 & 0.05671 & 0.04130 & \cellcolor[HTML]{B7E1CD}0.00002 & 0.04201 \\
\textbf{FC5} & \cellcolor[HTML]{B7E1CD}0.00009 & \cellcolor[HTML]{B7E1CD}0.00363 & \cellcolor[HTML]{B7E1CD}0.01120 & 0.25138 & 0.71212 & 0.03244 & 0.66335 & 0.18981 & 0.04338 & \cellcolor[HTML]{B7E1CD}0.00000 & \cellcolor[HTML]{B7E1CD}0.00012 \\
\textbf{FC1} & \cellcolor[HTML]{B7E1CD}0.02336 & 0.44807 & 0.74664 & 0.34585 & 0.75457 & 0.03649 & 0.54762 & 0.09789 & 0.53733 & 0.10794 & 0.08896 \\
\textbf{FC2} & 0.57941 & 0.04397 & 0.78418 & 0.26744 & 0.93032 & 0.36970 & \cellcolor[HTML]{B7E1CD}0.00059 & 0.79848 & 0.29156 & 0.32295 & 0.04679 \\
\textbf{FC6} & \cellcolor[HTML]{B7E1CD}0.00000 & 0.04885 & \cellcolor[HTML]{B7E1CD}0.01239 & 0.04684 & 0.04311 & \cellcolor[HTML]{B7E1CD}0.00014 & \cellcolor[HTML]{B7E1CD}0.00250 & 0.76418 & 0.04384 & \cellcolor[HTML]{B7E1CD}0.00007 & \cellcolor[HTML]{B7E1CD}0.00925 \\
\textbf{T7} & \cellcolor[HTML]{B7E1CD}0.00005 & 0.25188 & \cellcolor[HTML]{B7E1CD}0.01995 & 0.27151 & 0.20979 & \cellcolor[HTML]{B7E1CD}0.00607 & 0.79796 & 0.05030 & 0.07738 & \cellcolor[HTML]{B7E1CD}0.00191 & \cellcolor[HTML]{B7E1CD}0.00029 \\
\textbf{C3} & \cellcolor[HTML]{B7E1CD}0.00013 & \cellcolor[HTML]{B7E1CD}0.01320 & \cellcolor[HTML]{B7E1CD}0.01267 & \cellcolor[HTML]{B7E1CD}0.00341 & 0.15361 & \cellcolor[HTML]{B7E1CD}0.00126 & 0.41768 & 0.39888 & 0.23342 & 0.90493 & \cellcolor[HTML]{B7E1CD}0.00150 \\
\textbf{Cz} & 0.41892 & 0.47653 & 0.61352 & 0.47199 & 0.99340 & 0.54150 & 0.79046 & 0.14090 & 0.58900 & 0.30217 & 0.55975 \\
\textbf{C4} & \cellcolor[HTML]{B7E1CD}0.00002 & 0.15486 & \cellcolor[HTML]{B7E1CD}0.00801 & 0.21351 & 0.14819 & 0.76600 & 0.54410 & 0.49733 & 0.03597 & \cellcolor[HTML]{B7E1CD}0.01691 & 0.79268 \\
\textbf{T8} & \cellcolor[HTML]{B7E1CD}0.00015 & 0.16395 & 0.02600 & 0.28995 & 0.04407 & \cellcolor[HTML]{B7E1CD}0.01716 & 0.15327 & 0.20488 & 0.28142 & 0.05450 & \cellcolor[HTML]{B7E1CD}0.01530 \\
\textbf{CP5} & \cellcolor[HTML]{B7E1CD}0.00004 & 0.03887 & \cellcolor[HTML]{B7E1CD}0.00165 & 0.33954 & 0.48483 & \cellcolor[HTML]{B7E1CD}0.01427 & 0.41556 & 0.07253 & 0.12905 & 0.39827 & \cellcolor[HTML]{B7E1CD}0.00057 \\
\textbf{CP1} & 0.12915 & 0.44864 & 0.85573 & 0.24505 & 0.44346 & \cellcolor[HTML]{B7E1CD}0.01393 & 0.89985 & 0.42887 & 0.21951 & 0.51437 & 0.68123 \\
\textbf{CP2} & 0.33270 & 0.20764 & 0.96396 & 0.31162 & 0.55748 & 0.86790 & 0.81970 & 0.96763 & 0.55346 & 0.33392 & 0.27585 \\
\textbf{CP6} & \cellcolor[HTML]{B7E1CD}0.00000 & \cellcolor[HTML]{B7E1CD}0.00640 & \cellcolor[HTML]{B7E1CD}0.00004 & 0.29732 & \cellcolor[HTML]{B7E1CD}0.00004 & 0.65421 & 0.26598 & 0.34070 & \cellcolor[HTML]{B7E1CD}0.02197 & 0.36584 & 0.48809 \\
\textbf{P7} & 0.09157 & \cellcolor[HTML]{B7E1CD}0.00263 & \cellcolor[HTML]{B7E1CD}0.00021 & 0.28964 & 0.59462 & \cellcolor[HTML]{B7E1CD}0.00331 & 0.29223 & 0.29063 & 0.04477 & 0.34309 & \cellcolor[HTML]{B7E1CD}0.00001 \\
\textbf{P3} & 0.97236 & 0.23690 & 0.68281 & 0.16012 & 0.91648 & \cellcolor[HTML]{B7E1CD}0.02277 & 0.10400 & 0.22530 & 0.16049 & 0.50955 & 0.56544 \\
\textbf{Pz} & 0.85218 & 0.56220 & 0.09459 & 0.27451 & 0.90142 & 0.07456 & 0.10959 & 0.30934 & 0.41178 & 0.35498 & 0.63410 \\
\textbf{P4} & \cellcolor[HTML]{B7E1CD}0.00423 & 0.17090 & 0.17432 & 0.30756 & 0.25364 & 0.62829 & 0.75898 & 0.98663 & 0.03790 & 0.13737 & 0.79855 \\
\textbf{P8} & \cellcolor[HTML]{B7E1CD}0.00027 & \cellcolor[HTML]{B7E1CD}0.01414 & \cellcolor[HTML]{B7E1CD}0.01287 & 0.35839 & \cellcolor[HTML]{B7E1CD}0.00014 & 0.62500 & \cellcolor[HTML]{B7E1CD}0.01359 & 0.81521 & 0.12431 & 0.16817 & 0.94844 \\
\textbf{POz} & 0.89239 & 0.70536 & 0.54979 & 0.29913 & 0.48365 & 0.80864 & 0.32441 & 0.48797 & 0.37951 & \cellcolor[HTML]{B7E1CD}0.02381 & 0.36966 \\
\textbf{O1} & \cellcolor[HTML]{B7E1CD}0.01833 & 0.26995 & \cellcolor[HTML]{B7E1CD}0.00330 & 0.31165 & 0.41229 & 0.40539 & \cellcolor[HTML]{B7E1CD}0.02333 & 0.21469 & 0.12276 & 0.58767 & 0.03315 \\
\textbf{O2} & 0.04116 & 0.76373 & 0.90665 & 0.30193 & 0.11906 & 0.96514 & 0.03857 & 0.50735 & 0.02998 & 0.07279 & 0.46232 \\
\textbf{AF7} & \cellcolor[HTML]{B7E1CD}0.00003 & 0.10436 & 0.15344 & 0.30185 & \cellcolor[HTML]{B7E1CD}0.01205 & 0.04714 & 0.85248 & 0.17806 & 0.19249 & \cellcolor[HTML]{B7E1CD}0.00973 & \cellcolor[HTML]{B7E1CD}0.00120 \\
\textbf{AF3} & \cellcolor[HTML]{B7E1CD}0.02492 & 0.44451 & 0.10699 & 0.73860 & 0.51935 & 0.17558 & 0.20136 & 0.86716 & 0.38522 & 0.22531 & \cellcolor[HTML]{B7E1CD}0.00022 \\
\textbf{AF4} & 0.14204 & 0.02565 & 0.17741 & 0.29993 & 0.81433 & 0.20511 & 0.36660 & 0.90091 & 0.04390 & 0.28112 & \cellcolor[HTML]{B7E1CD}0.00103 \\
\textbf{AF8} & \cellcolor[HTML]{B7E1CD}0.02023 & \cellcolor[HTML]{B7E1CD}0.00016 & \cellcolor[HTML]{B7E1CD}0.00173 & 0.79399 & 0.31113 & 0.13581 & 0.88392 & 0.25631 & 0.05965 & 0.03151 & \cellcolor[HTML]{B7E1CD}0.01707 \\
\textbf{F5} & \cellcolor[HTML]{B7E1CD}0.00000 & 0.53116 & 0.45564 & 0.11112 & 0.82216 & \cellcolor[HTML]{B7E1CD}0.01562 & 0.58645 & \cellcolor[HTML]{B7E1CD}0.00006 & 0.32305 & \cellcolor[HTML]{B7E1CD}0.00000 & \cellcolor[HTML]{B7E1CD}0.00018 \\
\textbf{F1} & \cellcolor[HTML]{B7E1CD}0.01282 & 0.56591 & 0.11016 & \cellcolor[HTML]{B7E1CD}0.01448 & 0.23511 & 0.10754 & 0.92323 & 0.31363 & 0.38355 & 0.28459 & 0.02550 \\
\textbf{F2} & 0.17303 & 0.08604 & 0.25262 & 0.25653 & 0.59430 & 0.08602 & 0.48581 & 0.76021 & 0.13643 & 0.27295 & 0.02659 \\
\textbf{F6} & \cellcolor[HTML]{B7E1CD}0.00000 & 0.15170 & 0.02531 & 0.16655 & \cellcolor[HTML]{B7E1CD}0.01047 & \cellcolor[HTML]{B7E1CD}0.02109 & 0.11712 & 0.28348 & 0.03957 & \cellcolor[HTML]{B7E1CD}0.00084 & \cellcolor[HTML]{B7E1CD}0.01496 \\
\textbf{FC3} & \cellcolor[HTML]{B7E1CD}0.00027 & \cellcolor[HTML]{B7E1CD}0.01762 & 0.09605 & 0.23778 & 0.05743 & \cellcolor[HTML]{B7E1CD}0.00070 & 0.55411 & 0.16571 & 0.59173 & \cellcolor[HTML]{B7E1CD}0.00540 & \cellcolor[HTML]{B7E1CD}0.00304 \\
\textbf{FCz} & 0.04396 & 0.42688 & 0.90522 & 0.35778 & 0.93261 & 0.33650 & 0.55437 & 0.41474 & 0.59068 & 0.39384 & 0.11599 \\
\textbf{FC4} & \cellcolor[HTML]{B7E1CD}0.02457 & 0.05151 & \cellcolor[HTML]{B7E1CD}0.00004 & 0.12513 & 0.60100 & 0.43816 & \cellcolor[HTML]{B7E1CD}0.00225 & 0.02659 & 0.03685 & \cellcolor[HTML]{B7E1CD}0.00906 & 0.28673 \\
\textbf{C5} & \cellcolor[HTML]{B7E1CD}0.00003 & \cellcolor[HTML]{B7E1CD}0.01467 & 0.03565 & 0.26088 & 0.66043 & 0.08721 & 0.03954 & 0.22887 & 0.24832 & \cellcolor[HTML]{B7E1CD}0.00053 & \cellcolor[HTML]{B7E1CD}0.00028 \\
\textbf{C1} & \cellcolor[HTML]{B7E1CD}0.00001 & 0.04255 & 0.87720 & 0.41059 & 0.97966 & 0.06294 & 0.87731 & 0.11844 & 0.55636 & 0.10691 & 0.25221 \\
\textbf{C2} & 0.59287 & 0.34161 & 0.53022 & 0.33150 & 0.70296 & 0.81661 & 0.92870 & 0.94627 & 0.46839 & 0.26990 & 0.50454 \\
\textbf{C6} & \cellcolor[HTML]{B7E1CD}0.00000 & \cellcolor[HTML]{B7E1CD}0.00912 & 0.03691 & 0.44689 & 0.16790 & \cellcolor[HTML]{B7E1CD}0.02187 & 0.54138 & 0.24622 & 0.06441 & \cellcolor[HTML]{B7E1CD}0.00010 & 0.19130 \\
\textbf{CP3} & \cellcolor[HTML]{B7E1CD}0.00212 & 0.04331 & 0.08342 & 0.14651 & 0.43228 & \cellcolor[HTML]{B7E1CD}0.00579 & 0.86854 & 0.54992 & 0.20208 & 0.07623 & 0.26430 \\
\textbf{CP4} & 0.05828 & 0.07029 & 0.06293 & 0.32495 & 0.41968 & 0.96079 & 0.87827 & 0.50303 & 0.22450 & 0.22669 & 0.76220 \\
\textbf{P5} & 0.11911 & \cellcolor[HTML]{B7E1CD}0.00005 & 0.64300 & 0.28803 & 0.60948 & \cellcolor[HTML]{B7E1CD}0.01478 & 0.42166 & 0.05911 & 0.06916 & 0.52824 & \cellcolor[HTML]{B7E1CD}0.00025 \\
\textbf{P1} & 0.06312 & 0.34918 & 0.18271 & 0.26193 & 0.99534 & 0.03671 & 0.17718 & 0.43583 & 0.29674 & \cellcolor[HTML]{B7E1CD}0.01134 & 0.95687 \\
\textbf{P2} & 0.91909 & 0.38238 & 0.64611 & 0.10739 & 0.83258 & 0.77928 & 0.37719 & 0.15761 & 0.45133 & 0.11639 & 0.74606 \\
\textbf{P6} & \cellcolor[HTML]{B7E1CD}0.00023 & \cellcolor[HTML]{B7E1CD}0.00027 & \cellcolor[HTML]{B7E1CD}0.00000 & 0.33048 & \cellcolor[HTML]{B7E1CD}0.00011 & 0.80477 & 0.33497 & 0.63682 & 0.05144 & 0.03666 & 0.93989 \\
\textbf{PO5} & 0.04774 & 0.44393 & 0.75352 & 0.32452 & 0.51866 & 0.54693 & 0.04441 & 0.17626 & 0.13882 & 0.50722 & 0.03686 \\
\textbf{PO3} & 0.08718 & 0.48565 & 0.90504 & 0.30817 & 0.65918 & 0.47752 & 0.15085 & 0.17811 & 0.14202 & 0.51446 & 0.08102 \\
\textbf{PO4} & \cellcolor[HTML]{B7E1CD}0.00643 & 0.45273 & 0.02624 & 0.32006 & 0.15017 & 0.84428 & 0.31105 & 0.82650 & 0.03128 & 0.12982 & 0.77907 \\
\textbf{PO6} & \cellcolor[HTML]{B7E1CD}0.00710 & 0.10452 & 0.05401 & 0.31456 & 0.12165 & 0.79154 & 0.23823 & 0.76810 & 0.03782 & 0.15941 & 0.10852 \\
\textbf{FT7} & \cellcolor[HTML]{B7E1CD}0.00007 & 0.02902 & 0.07481 & 0.32132 & 0.09018 & 0.02658 & 0.57583 & 0.45973 & 0.03600 & \cellcolor[HTML]{B7E1CD}0.00877 & \cellcolor[HTML]{B7E1CD}0.00016 \\
\textbf{FT8} & \cellcolor[HTML]{B7E1CD}0.00001 & 0.94476 & 0.47023 & 0.60191 & 0.04755 & 0.09358 & 0.64506 & 0.04385 & 0.03129 & \cellcolor[HTML]{B7E1CD}0.00246 & \cellcolor[HTML]{B7E1CD}0.00945 \\
\textbf{TP7} & \cellcolor[HTML]{B7E1CD}0.00170 & \cellcolor[HTML]{B7E1CD}0.00036 & 0.07220 & 0.16509 & 0.05615 & 0.06854 & 0.16858 & 0.34853 & 0.04948 & \cellcolor[HTML]{B7E1CD}0.00360 & \cellcolor[HTML]{B7E1CD}0.00704 \\
\textbf{TP8} & \cellcolor[HTML]{B7E1CD}0.00004 & \cellcolor[HTML]{B7E1CD}0.02309 & \cellcolor[HTML]{B7E1CD}0.00009 & 0.44326 & \cellcolor[HTML]{B7E1CD}0.01359 & 0.07991 & 0.18782 & 0.16156 & 0.05211 & 0.14219 & 0.62296 \\
\textbf{PO7} & 0.05902 & 0.43647 & 0.65610 & 0.30304 & 0.49801 & 0.45075 & 0.03270 & 0.26861 & 0.13704 & 0.50959 & 0.02852 \\
\textbf{PO8} & \cellcolor[HTML]{B7E1CD}0.00785 & 0.83909 & 0.13406 & 0.31248 & 0.11909 & 0.80972 & 0.28353 & 0.75029 & 0.03654 & 0.12702 & 0.14667 \\
\textbf{Oz} & 0.33123 & 0.93008 & 0.96570 & 0.30176 & 0.50509 & 0.87993 & 0.10595 & \cellcolor[HTML]{B7E1CD}0.02219 & 0.35471 & 0.21248 & 0.12421
\end{tabular}
\end{table}

\clearpage
\section{Ensembling Accuracy}

\begin{table}[h]
\centering
\caption{\textit{Performance Summary of Ensemble Method per Participant}}
\label{app:ensembling-accuracy}
\begin{tabular}{l|rrlll}
\hline
\multicolumn{1}{c}{\textbf{Participant}} &
  \multicolumn{1}{c}{\textbf{Ensemble Training}} &
  \multicolumn{1}{c}{\textbf{Ensemble Testing}} & \\
\hline
p01 & 91.22\% & 82.55\% & \\
p02 & 67.58\% & 64.77\% & \\
p03 & 88.25\% & 81.33\% & \\
p04 & 69.64\% & 71.31\% & \\
p05 & 91.44\% & 80.22\% & \\
p06 & 96.73\% & 85.15\% & \\
p07 & 85.83\% & 79.11\% & \\
p08 & 91.30\% & 89.55\% & \\
p09 & 68.77\% & 67.55\% & \\
p10 & 87.55\% & 82.55\% & \\
p11 & 60.61\% & 66.11\% & \\
\hline
\end{tabular}
\end{table}

\section{Eye Tracking Classification}
\label{app:eye-tracking-classification}

\begin{table}[h]
\caption{\textit{Eye Tracking CNN Architecture.} Input data (2-channel time-series data containing x and y gaze coordinates) pass through three convolution, pooling, batch normalization, and activation layers, followed by two linear layers with GELU activation.}
\label{tab:tab_g1}
\centering
\begin{tabular}{l|ll}
\hline
Layer (type)   & Output Shape       & Param \# \\
\hline
Conv1d-1       & {[}-1, 16, 1000{]} & 176      \\
BatchNorm1d-2  & {[}-1, 16, 1000{]} & 32       \\
ReLU-3         & {[}-1, 16, 1000{]} & 0        \\
AvgPool1d-4    & {[}-1, 16, 500{]}  & 0        \\
Conv1d-5       & {[}-1, 32, 500{]}  & 2592     \\
BatchNorm1d-6  & {[}-1, 32, 500{]}  & 64       \\
ReLU-7         & {[}-1, 32, 500{]}  & 0        \\
AvgPool1d-8    & {[}-1, 32, 250{]}  & 0        \\
Conv1d-9       & {[}-1, 64, 250{]}  & 10304    \\
BatchNorm1d-10 & {[}-1, 64, 250{]}  & 128      \\
ReLU-11        & {[}-1, 64, 250{]}  & 0        \\
AvgPool1d-12   & {[}-1, 64, 125{]}  & 0        \\
Linear-13      & {[}-1, 128{]}      & 1024128  \\
GELU-14        & {[}-1, 128{]}      & 0        \\
Linear-15      & {[}-1, 1{]}        & 129      \\
\hline
\end{tabular}
\end{table}

\end{document}